\documentclass[12pt]{article}
\usepackage{amsfonts}
\usepackage{amssymb}

\input{tcilatex}
\begin{document}

\begin{center}
{\large The Sound of Silence: equilibrium filtering and optimal censoring in
financial markets}

by

Miles B. Gietzmann and Adam J. Ostaszewski

\bigskip

\textsc{arXiv version}

\bigskip
\end{center}

\textbf{Abstract. }Following the approach of standard filtering theory, we
analyse investor-valuation of firms, when these are modelled as
geometric-Brownian state processes that are privately and partially
observed, at random (Poisson) times, by agents. Tasked with disclosing
forecast values, agents are able purposefully to withhold their
observations; explicit filtering formulas are derived for downgrading the
valuations in the absence of disclosures. The analysis is conducted for both
a solitary firm and $m$ co-dependent firms.


\textbf{AMS\ Classification: 91G50, 91G80; 93E11, 93E35; 60G35, 60G25.}

\section{Introduction}

Motivated by applications from financial mathematics, we assume below that
the flow of information reaching a market is influenced strategically by the
agents responsible for making and subsequently disclosing observations,
simply because observations made by a self-interested agent can be
suppressed. For instance, suppose the agent's self-interest manifests itself
as wanting only to report good news, and so to suppress bad news. If the
cutoff that defines whether an observation is good or bad news is defined
relative to prior expectations, then the agent will disclose only
observations above the prior. A model of strategic reporting needs
specifically to incorporate how potential recipients of disclosed
observations (investors) can rationally respond to \textit{silence} from the
agent, when absence of observation cannot credibly be indicated by the agent
and disclosure is not mandatory. We assume below that at Poisson arrival
times the agent is able to observe information, and has then the option to
disclose the observed information, but only truthfully, or to suppress it
(i.e. hide the `bad' news). Then the question arises: what is the optimal
level of suppression. We answer below in Theorem 1 and Theorem 1$_{m}$ with
explicit formulas in a multi-agent multi-period context. We work in a
continuous Black-Scholes framework introducing the novel concept of an 
\textit{optimal censor }(\S 2.2). This combines censoring with filtering
(for which see \cite{AggE}, \cite{BaiC}, \cite{Dav1}, \cite{LipS}). It is
the key here. In particular we derive closed-form \textit{exponential decay}
solutions for the optimal time-varying censor in periods of silence; a new
insight here is that penalization of silence is harshest at the beginning
(see \S 3.2 Corollary for details). These are the pleasing consequence of
moving to continuous time: simplification of the natural equilibrium
conditions (e.g. (\ref{Dye-1}) in \S 2.3 and in \S 3.1.1; cf. (\ref{NE})) to
a first-order differential equation (see (\ref{basic-de}) in \S 3.1); a
transparent narrative at the single agent level; a rich joint asset-price
dynamic, via the repercussions on each other, of a stream of disclosures
from the multiplicity of correlated agents. This paper is a sequel to our
previous work, where such questions were pursued in a discrete two-period
setting (in \cite{OstG} and \cite{GieOb} for the case of one firm, and in 
\cite{GieOa} where further reality is added via a communication game with
multiple competitors in an industry correlated by common operating
conditions). We were motivated by the static-setting literature of costly
state verification (e.g. by Townsend \cite{Tow} in 1979 -- see the later
literature in \cite{KraV}), and of corporate disclosure introduced by Dye 
\cite{Dye} in 1985 (and the associated paper \cite{JunK}).

The additional results in \S 4 highlight consequences of our main theorems
for the bandwagon and quality effects in the current setting. These
qualitative results extend our findings in the two-period model of \cite%
{GieOa}. For instance: when competing managers are endowed with different
observation noise, those managers that observe with most noise use a lower
censor and hence suppress bad news less. But then this means that, when
investors see \textit{slightly-below-mean} observations being disclosed by
such managers, they rationally interpret this as being from a more noisy
source, and so discount its importance when updating.

A broadly similar class of models arises in the engineering literature
studying alternation between observation (`measurement') and control,
typically in a discrete-time setting, but there the alternation is the
result of a trade-off between the two actions, dictated by a \textit{single}
indicator of overall performance of a system (i.e. a suitable objective
function); a related class considers intermittent receipt of
measurements/observations, but there the suppression is caused by random
transmission failures -- see e.g. \cite{CalN}, \cite{ImmB}, \cite{SinAl}.

In discrete time, Shin \cite{Shi} introduced (in 2004) a model of a firm,
engaged in a flow of projects, receiving information with a Poisson
distribution concerning the status (success or failure) of projects
completed to date. By endowing the firm with the opportunity at each date of
a full or partial disclosure of that information, he created an
asset-pricing framework in which such disclosures are endogenously
determined, and so could study equilibrium patterns of corporate disclosure.

In continuous time, Brody, Hughston and Macrina \cite{BroHM} introduced (in
2007) an asset-pricing framework based on noisy market-observation of
continuous information that is generated from the remaining future
disclosures, occuring at pre-determined (`mandatory') dates. Similarly to
Shin, their disclosures are modelled as a discrete series of random
variables; the latter give rise to a stream of uncertain payouts (cash
flows) corresponding to the mandatory dates. The individual cash values are
taken to be deterministic functions of independent random variables called
`market-factors' (not unlike our terminal-time output variable $Z_{1}$), but
with the inclusion only of those that are capable of being observed by the
modeller at and prior to the date of disclosure. Typically the market's
noisy observation is a linear combination of the future payouts, each
weighted by a coefficient that gains increasing prominence over time, and
Brownian-bridge components, which vanish at the corresponding payout-dates.
However, in contrast to Shin, all the disclosures are mandatory and the
possibility of voluntary intermittent disclosures (say via a zero dividend)
is not studied.

In a recent development, Marinovic and Varas \cite{MarV} (in 2014) consider
a voluntary disclosure model in continuous time, in which a single agent
uninterruptedly observes an asset following a $\{0,1\}$-valued random walk.
(The binary aspect leads to market-price decay, which like ours is
exponential.)

In creating a multi-asset Black-Scholes asset-pricing framework in which
corporate disclosures are endogenously determined, our work is closest in
spirit to Shin \cite{Shi}. It is also closer in spirit with some of the
earlier literature of portfolio/consumption analysis under incomplete
information, e.g. Feldman's model\ in 1992 (\cite{Fel}; cf. literature cited
there) of a production exchange economy, where realized outputs are observed
(whereas in the model below a noisy version of the output process $Z_{t}$ is
observed), and provide via a nonlinear filter an information flow on the
underlying \textit{economic state process} (a `productivity' factor,
following an Ornstein--Uhlenbeck mean-reverting process).

The rest of this paper is structured as follows. In \textbf{\S }2,
abstracting away from the market motivation, we exploit in \S 2.1 some of
the ideas of standard filtering theory to describe scalar observations by $m$
individuals (`agents') that are made intermittently in the interval $(0,1)$
privately (i.e. in secret), and either kept secret or disclosed (reported)
to the other agents and investors. The observation processes have a
co-dependance, as do also certain other scalar processes which assign a
valuation to each agent. This co-dependance is determined by a single
state-process called the \textit{common effect}. The shared information,
together with the initial mandatory observation values of time $t=0,$
provides the basis upon which to forecast each individual's valuation at the
terminal time $t=1$ of the next mandatory disclosure.

Agents disclose their observations in order to enhance at each moment the
forecast of their terminal valuation, which is contingent on the mandatory
disclosures at the terminal time. This leads to an optimization problem
stated in \textbf{\S }2.2. If each agent discloses only observations above
their \textit{censoring} threshold, keeping secret lower observed values,
silence at any moment in $(0,1)$ can mean either the absence of observation,
or its censoring (since the arrival of an undisclosed observation remains
secret, and the agent is not able to assert credibly the absence of a
current observation). Since periods of silence bring precautionary valuation
downgrades, the threshold drops over time, pari passu, to elicit a
disclosure; additional tensions arise from potential disclosures from other
agents, so subgame perfect Nash equilibria considerations enter the argument
-- see \S 3.

Section 2.3 interprets the individual valuation as the market value of the
firm obtained from information disclosed to the market by managers making
discretionary (i.e. non-mandatory) disclosures in between the two mandatory
disclosure times of $t=0$ and $t=1.$

Our two main results are stated and discussed in \textbf{\S }3 and proved in 
\textbf{\S }5.2, \S 5.3. In \textbf{\S }4 the interpretation of \textbf{\S }%
2.3 is used to describe, as immediate corollaries of earlier work,
qualitative features of a multi-agent correlated sector.

We focus on valuations inferred from periods of silence hence the title.

\section{Model}

We first formalize the disclosure framework in a series of steps.

\subsection{Processes and filtrations}

We fix a stochastic basis $(\Omega ,\mathcal{F},\mathbb{F},\mathbb{Q})$,
with $\mathbb{F}:=\{\mathcal{F}_{t}:t\in \lbrack 0,1]\}$. \medskip

\textit{2.1.1. State process setup.}\enspace The production of $m\geq 1$
scalar processes from one (vector) process, and the uncertainties connected
with this, will be modelled in terms of the fixed $(\mathbb{F}$-$\mathbb{Q})$%
-Wiener process $W_{t}=(W_{t}^{0},W_{t}^{1},...,W_{t}^{m})$, more precisely
in terms of exponential martingales $X$ and $M^{i}$ defined from the
independent component processes of $W,$ employing correspondingly the
entries of the real vector $\sigma =(\sigma _{0}^{X},\sigma _{1}^{M},\ldots
,\sigma _{m}^{M})\geq 0$ as follows\footnote{%
Our choice of working with exponential martingales, indeed with Dol\'{e}%
ans-Dade exponentials, is natural in a setting where a critical equation
involves equity values rather than returns: see Prop. 1 below and especially
the equity valuation equation (\ref{risk-neutral}), as well as the interplay
of the key equations: (\ref{Dye}), (\ref{JungKwon7}) [equivalently, (\ref%
{H-LN})], and likewise (\ref{NE}).}: 
\begin{eqnarray*}
dX_{t}/X_{t} &=&\sigma _{0}^{X}\,dW_{t}^{0}\,,\quad t\in \lbrack 0,1]\,, \\
dM_{t}^{i}/M_{t} &=&\sigma _{i}^{M}\,dW_{t}^{i}\,,\quad t\in \lbrack 0,1]\,,
\end{eqnarray*}%
for every $i$ $\in I:=\{1,\ldots ,m\}$; here we assume as known the
respective distributions at time-$0$ of these processes.

The process $X$ has the dual roles of \textit{(economic) state} process and 
\textit{signal} process, and is regarded as representing a \textit{common
effect} that influences the individual internal histories via two processes
as follows.

\textit{2.1.2 Ouput process.} The \textit{output process} $%
Z_{t}=(Z_{t}^{1},...,Z_{t}^{m})$, whose terminal state at time $t=1$ is to
be forecast, encodes the production of $m$ signals from the state process as
its single input; taking values in $\mathbb{R}^{m}$, it is defined
componentwise by a weighted power-law: 
\[
Z_{t}^{i}=\zeta ^{i}(X_{t}),\quad \mathrm{where}\quad \zeta
^{i}(x)=f^{i}x^{\alpha _{i}}\,, 
\]%
for $i\in \{1,\ldots ,m\}$; here $f^{i}$ is the \textit{size constant} of
the $i$-th output and $\alpha _{i}$ is termed the $i$-th \textit{loading
exponent (loading factor)}, both positive, yielding the \textit{terminal
output} vector $Z_{1}$. The exponent, which measures the exposure of a firm
to common effects, enables empirical interpretation but is mathematically of
little consequence (as though it had the value 1). To add a technical
remark, it will be convenient in the proofs to rescale such a process to
have size unity at some date, contingent on the available information, a
procedure we call \textit{common-sizing}.

\textit{2.1.3 Internal history process.} The \textit{internal history process%
} $Y_{t}=(Y_{t}^{1},\ldots ,Y_{t}^{m})$ provides the source of intermittent
noisy observations (which are to be selectively disclosed) of the output
process $Z_{t}$; as this is to be a partially observable process, the
construction uses the martingale $M=(M^{1},\ldots ,M^{m})$ as a linear noise
to modify the output process: 
\[
Y_{t}=Z_{t}M_{t}, 
\]%
that is, componentwise: $Y_{t}^{i}=Z_{t}^{i}M_{t}^{i}=f^{i}X_{t}^{\alpha
_{i}}M_{t}^{i}.$

\textit{2.1.4\enspace Regression parameters.}\enspace We collect regression
parameters associated with the processes defined so far. Put $\sigma
_{i}:=\sigma _{i}^{M}/\alpha _{i}$, and define the associated \textit{%
precisions} 
\[
p_{i}:=1/\sigma _{i}^{2}\text{ and }p=p_{\text{agg}}:=p_{0}+p_{1}+...+p_{m}; 
\]%
for convenience, we use dual notation for the regression coefficients
(relative precisions): 
\[
\kappa _{i}\text{ or }\kappa _{m}^{i}:=p_{i}/p\text{ and }\kappa
_{1}^{i}=p_{i}/(p_{0}+p_{i})\,\text{for }m=1\text{ with }i\text{ the
solitary firm.} 
\]

\textit{2.1.5 Private observation process and private filtration}. The
observation of the economic realities embodied in the processes $X$, $Y$,
and $Z$ is modelled as an \textit{intermittent partial} \textit{observation}
arising for an individual economic agent with label in $\{1,\ldots ,m\}$,
and expressed as a \textit{private observation process} $Y^{i\text{-obs}}$.
The constructions of these concepts below starts from a (c\`{a}dl\`{a}g)
Poisson process $N_{t}=(N_{t}^{1},...,N_{t}^{m})$ which is independent of
the Wiener processes $W_{t}$ of \S 2.2.1 and whose vector of intensity
functions $\lambda _{t}=(\lambda _{t}^{1},\ldots ,\lambda _{t}^{m}),$ the 
\textit{information arrival rate}, is inhomogeneous and also known. The
arrival times in $(0,1)$ of each component process $N^{i}$ are regarded as
consecutively numbered, with $\theta _{n}^{i\text{-obs}}$ denoting the $n$%
-th of these; that is, on setting $\theta _{0}^{i\text{-obs}}=0,$ 
\[
\theta _{n}^{i\text{-obs}}:=\inf \{t>\theta _{n-1}^{i\text{-obs}%
}:N^{i}(t)>N^{i}(t-)\}. 
\]%
The intended meaning of this is that the $i$-th agent observes the process $%
Y^{i\text{-obs}}$ privately, and only at these arrival time in $(0,1)$; the
last observation time at or prior to $t$ and the corresponding last observed
value are 
\begin{eqnarray*}
\theta _{-}^{i\text{-obs}}(t) &=&\max \{s\leq t:N^{i}(s)>N^{i}(s-)\}, \\
Y^{i\text{-obs}}(t) &=&Y^{i}(\theta _{-}^{i\text{-obs}}(t)).
\end{eqnarray*}%
The resulting process $Y^{i\text{-obs}}$ is the \textit{private observation
process} of agent $i$. It is piecewise constant, and defines the \textit{%
private} \textit{filtration} of the $i$-th agent. This is the (time indexed)
family $\mathbb{Y}_{i}^{\text{priv}}$ of $\sigma $-algebras generated by the
jumps at or before time $t$, and by the observations at or before time $t,$
here regarded as \textit{space-time }point-processes\textit{\ }(i.e. with
their dates -- for background see \cite[esp. II Ch. 15]{DalVJ} ); it is
formally given by 
\[
\mathbb{Y}_{i}^{\text{priv}}:=\{\mathcal{Y}_{t}^{i}:t\in \lbrack 0,1]\},%
\text{ where }\mathcal{Y}_{t}^{i}:=\sigma (\{(s,Y^{i\text{-obs}%
}(s),N^{i}(s)):0<s\leq t\})\,, 
\]%
for $t\in (0,1),$ where we let $\mathcal{Y}_{0}^{i}$ contain all the null
sets of $\mathcal{F}$.

\textit{2.1.6 Public filtration}. The \textit{public filtrations \/}$\mathbb{%
G}^{\text{pub}}$, to be constructed next, formalize the notion of
`information disclosed via the publicly observed history of $Y$'. Their
construction begins with a fixed marked point-process\footnote{%
As the Referee points out, it possible to construct the public filtration
entirely from a vector of Poisson processes (with appropriate thinning), so
that `disclosers' are identified by coincidence of certain arrival times.}
(MPP) comprising the following list of items (i) to (iv).

\bigskip

\noindent (i) The underlying c\`{a}dl\`{a}g counting process $N_{t}^{\text{%
pub}}$ (`the disclosure time process'),

\noindent (ii) The functions $\theta _{\pm }^{\text{pub}}$, where $\theta
_{-}^{\text{pub}}(t)$ is the last arrival time of $N_{t}^{\text{pub}}$ less
than or equal to $t$, and $\theta _{+}^{\text{pub}}(t)$ is the first arrival
time of $N_{t}^{\text{pub}}$ bigger than or equal to $t$, these relations
holding almost surely on $\mathcal{F}_{t}$, for every $t$ in $[0,1]$.

\noindent (iii) The point process $J_{t}\subseteq I:=\{1,\ldots ,m\}$ (`the
disclosing agent set of time $t$').

\noindent (iv) The marks $Y_{t}^{j\text{-pub}}=Y^{j\text{-obs}}(\theta _{-}^{%
\text{pub}}(t))$ for $j\in J_{t}$ (`their corresponding observations of time 
$t$').

\bigskip

In terms of the MPP as above we obtain the \textit{publicly observed history}%
, or \textit{public filtration}, which is right-continuous and constructed
in the three steps below: 
\[
\mathbb{G}=\mathbb{G}^{\text{pub}}:=\{\mathcal{G}_{t}^{+}:t\in \lbrack
0,1]\}\,, 
\]%
where $\mathcal{G}_{1}^{+}=\sigma (\mathcal{G}_{1},\{(1,Y^{i}(1))\}_{i\in
I}) $ and for each $t$ in $[0,1)$, conventionally 
\[
\mathcal{G}_{t}^{+}=\bigcap\nolimits_{s>t}\mathcal{G}_{s}, 
\]%
with the $\sigma $-algebras $\mathcal{G}_{t}$ generated as the join of the
three $\sigma $-algebras corresponding to the items (i), (iii) and (iv) from 
\S 2.1.6, dates included, namely:%
\[
\mathcal{G}_{t}=\vee _{s\in \lbrack 0,t)}\sigma \big((s,N^{\text{pub}%
}(s)),J_{s},\{(s,Y^{j\text{-pub}}(s))\}_{j\in J_{s}}\big). 
\]%
\smallskip \textit{2.1.7 Disclosure filtration}. The key concept for the
paper is that of a public filtration $\mathbb{G}$ (in the sense of \S 2.1.6)
which is a particular kind of subfiltration of the join $\vee _{i}\mathbb{Y}%
_{i}^{\text{priv}}$ of the private filtrations of \S 2.1.5. Its definition
hinges on disclosure of observations at or above the current value of a
`reference process' $\gamma $ which, \textit{crucially,} is required to be
predictable with respect to the public information that it itself generates.

We define the \textit{disclosure filtrations} \textit{consistently generated}
\textit{from} the (private) filtrations $\{\mathbb{Y}_{i}^{\text{priv}}:i\in
I\}$ \textit{via} the $\mathbb{G}$-predictable \textit{censoring filter }$%
\gamma =(\gamma _{t}^{i}:i\in I)$, to be public filtrations satisfying
additionally the following three conditions for each $0<t<1$. These say that
at each disclosure time there are disclosing agents as in (v), with their
observations made public, as in (vi), because, as in (vii), these are above
their censoring thresholds:\newline
\newline
\noindent (v) The set $J_{t}$ is non-empty iff $N^{\text{pub}}(t)>N^{\text{%
pub}}(t-)$.

\noindent (vi) If $N^{\text{pub}}(t)>N^{\text{pub}}(t-),$ then $Y^{j\text{%
-pub}}(t)=Y^{j\text{-obs}}(t),$ for each $j\in J_{t}$.

\noindent (vii) If $N^{i}(t)>N^{i}(t-)$ and $Y_{t}^{i\text{-obs}}>\gamma
_{t}^{i}$, for some $i\in I,$

\qquad then $N^{\text{pub}}(t)>N^{\text{pub}}(t-)$ and $i\in J_{t}$.

\bigskip

For such a filtration $\mathbb{G}^{\text{pub}}$, the counting-process
arrival times occuring in $(0,1)$ in (ii) of \S 2.1.6 will be termed the 
\textit{voluntary disclosure event times} (or just \textit{disclosure times}
if $m=1$): $\theta _{0}^{\text{pub}},\theta _{1}^{\text{pub}},...$ .

\subsection{Optimal censoring problem}

Associated with a fixed disclosure filtration $\mathbb{G}=\{\mathcal{G}%
_{t}^{+}\}$, consistently generated via $\gamma $ as in \S 2.1.7, is the
process obtained by taking contingent expectations of the time-$1$ output $%
Z_{1}$ with respect to the time-$t$ information subsets: 
\[
t\mapsto \mathbb{E}[Z_{1}^{i}|\mathcal{G}_{t}]\,; 
\]%
this process is interpreted as a $\mathbb{G}$-\textit{predictable valuation
process} or $\mathbb{G}$-\textit{forecasting process}. The \textit{optimal
censoring problem} addressed below calls for the construction of a
filtration $\mathbb{G}_{\ast }$, necessarily unique, whose associated
forecasting process is the \textit{left-sided-in-time} pointwise supremum
over all $\mathbb{G}$-forecasting processes, that is, for each time $t\in
(0,1)$ and each agent $i,$ 
\begin{equation}
\mathbb{E}[Z_{1}^{i}|\mathcal{G}_{\ast ,t}]\,=\sup\nolimits_{\mathbb{G}}%
\mathbb{E}[Z_{1}^{i}|\mathcal{G}_{t}],  \tag{OC}
\end{equation}%
where the supremum ranges over the disclosure filtrations $\mathbb{G}=\{%
\mathcal{G}_{t}^{+}\}$ of \S 2.1.7. In (OC) both sides of the equation
depend only the public information available to the \textit{left} of the date%
\footnote{%
It is highly significant to the analysis that the date $t$ is inferable from
the conditioning $\sigma $-algebra $\mathcal{G}_{t}$ (likewise for $\mathcal{%
Y}_{t}^{i}$), hence its inclusion in the construction.} $t.$ By definition
such a $\mathbb{G}_{\ast }$, if it exists, is unique. The main results of
the paper give the solution in \textbf{\S }3 of the optimal censoring
problem for the geometric Brownian signal processes $X$ of \textbf{\S }2.1;
Theorem~1 corresponds to the one-dimensional case $m=1$, Theorem~1$_{m}$ to
the case of arbitrary integer dimension $m\geq 1$.

\noindent \textbf{Remarks 1.} Informally, the censoring problem requires
agent $i$ to reach disclosure/suppression decisions using only the history
of all prior public information.

\noindent \textbf{2.} With the filtering theory language above, the optimal
censoring problem amounts to the construction of a censoring-filter
(process) $\gamma _{t}$ with the following properties (i) to (iii):\newline
\noindent (i) the disclosure subfiltration $\mathbb{G}$ of $\vee _{i\in I}%
\mathbb{Y}_{i}^{\text{priv}}$, generated through suppression of observations
by reference to the process $\gamma ,$ is consistently generated from the
private filtrations $\{\mathbb{Y}^{i}:i=1,...,m\};$\newline
\noindent (ii) $\gamma $ is $\mathbb{G}$-predictable (this \textit{is}
crucial);\newline
\noindent (iii) for each $i\in I$ and for each time instant $t\in (0,1)$, $%
\gamma _{t}^{i}$ is chosen to be a cutoff \textit{maximizing} the expected
value of $Z_{1}^{i}$, given only past public information, \textit{and} best
response to the simultaneous \textit{censoring choices} of $\gamma _{t}^{j}$
for $j\neq i$ (which is why $\gamma $ needs to be $\mathbb{G}$-predictable).
Of course, this yields an individual \textit{private valuation }for each
agent\textit{.}

\noindent \textbf{3.} The role of $\mathbb{G}$ above is to formalize the
censoring of the private observation processes relative to information
`public' \textit{before} any time-$t$ disclosure, so in general distinct
from the \textit{optional valuation process} of the next section (\S 2.3)%
\[
\mathbb{E}[Z_{1}^{i}|\mathcal{G}_{t}^{+}], 
\]%
which models the later right-continuous \textit{public valuation} at time $t$%
.

\noindent \textbf{4.} In Remark 2 the agent is a maximizer of an
instantaneous objective linked to the terminal output (via its estimator).
The alternative approach is to establish a single overall performance
indicator for the entire trajectory of the estimator. Optimality of overall
economic behaviour induced by instantaneous (sometimes called `myopic')
objectives is established for a class of models related to ours in Feldman 
\cite{Fel}.

\subsection{Application to asset-price modelling}

We consider a single `firm' $Z_{t}^{i}$ in isolation, so omitting $i$ when
convenient. The manager of the firm (agent $i$) makes a mandatory
declaration at time $t=1$ of its (fundamental) value which, taking a
Bayesian stance, is the manager's estimate/forecast of the firm's economic
state $Z_{1}$ using $Y_{1},$ namely%
\[
\tilde{\gamma}_{1}:=\mathbb{E[}Z_{1}|\mathcal{G}_{1},Y_{1}]=\mathbb{E[}Z_{1}|%
\mathcal{G}_{1}^{+}], 
\]%
since $\mathcal{G}_{1}^{+}=\sigma ((1,Y_{1}),\mathcal{G}_{1})$. Below we use
a standard martingale construction to create an asset-price process under $%
\mathbb{Q}$ (cf. \cite{BinK}); we then note the option values that censoring
introduces, and observe that a censor, which induces indifference between
disclosure and non-disclosure, preserves the risk-neutral character of the
asset-price under $\mathbb{Q}$.

Given a public filtration $\mathbb{G}$, the associated forecasting process
of \textbf{\S }2.2. yields an analogue $S$ of an asset-price process with 
\[
S_{t}:=\mathbb{E}[\mathbb{E[}Z_{1}|\mathcal{G}_{1}^{+}]|\mathcal{G}%
_{t}^{+}],\quad \text{ for }0\leq t\leq 1. 
\]

This construction also turns the reference measure $\mathbb{Q}$ (of the
stochastic basis) into a risk-neutral valuation measure, a fact implied by
the conditional mean formula, which for $t<s$ asserts that 
\[
\mathbb{E}[S_{s}|\mathcal{G}_{t}^{+}]=\mathbb{E}[\mathbb{E}[\mathbb{E[}Z_{1}|%
\mathcal{G}_{1}^{+}]|\mathcal{G}_{s}^{+}]|\mathcal{G}_{t}^{+}]=\mathbb{E}[%
\mathbb{E[}Z_{1}|\mathcal{G}_{1}^{+}]|\mathcal{G}_{t}^{+}]=S_{t}. 
\]%
The disclosure cutoff $\gamma $ is uniquely determined below by the
asset-price process $S,$ as follows. Fix $t<s.$ At time $t$ suppose that
agent $i$ is committed to using a cutoff $\gamma _{t}$ and is to choose a
disclosure cutoff $\gamma $ for use at the later date $s$. Assume, until
further notice, absence of any public disclosures in the interval $(t,s).$
Let $D_{s}(\gamma )$ be the disclosure event of time $s$ corresponding to
the agent $i$ observing a value of $Y_{s}$ at or above $\gamma $, so that in
terms of indicator functions%
\begin{equation}
\mathbf{1}_{D_{s}(\gamma )}=\mathbf{1}_{N(s)>N(s-)}\cdot \mathbf{1}%
_{Y(s)\geq \gamma }.  \tag{D}
\end{equation}%
Equation (D) suggests that the Black-Scholes value of the \textit{%
one-or-nothing binary} \textit{option on }$Y_{s}$ with strike $\gamma $
(represented by $\mathbf{1}_{Y(s)\geq \gamma }$) will emerge at the heart of
our line of reasoning.

Next consider the complementary event $ND_{s}(\gamma )$ in which, given $%
\gamma ,$ the market computes a forecast for $Z_{1}$ as being $\tilde{\gamma}%
_{s}=\mathbb{E}[Z_{1}|ND_{s}(\gamma ),\mathcal{G}_{t}^{+}].$ If the agent
wishes to maximise the asset valuation $S$ at time $s$ the choice $\gamma $
will induce indifference between disclosure and non-disclosure when $%
Y_{s}=\gamma $ is observed iff\footnote{%
For simplicity, here and below we adopt the \textit{equational convention}
that conditioning on an equation $Y_{t}=y$ is to be read as implying its
disclosure.}%
\[
\mathbb{E}[Z_{1}|Y_{s}=\gamma ,\mathcal{G}_{t}^{+}]=\mathbb{E}%
[Z_{1}|ND_{s}(\gamma ),\mathcal{G}_{t}^{+}]. 
\]%
This observation and conditioning on knowledge at time $t$ of the value%
\footnote{%
This equals $\mathbb{E}[Z_{1}|Y_{t}=\gamma _{t},\mathcal{G}_{t}]$, absent
any disclosure.}%
\[
\tilde{\gamma}_{t}:=\mathbb{E}[Z_{1}|\mathcal{G}_{t}^{+}] 
\]%
induces the following simple relation between the equilibrium value $\tilde{%
\gamma}_{s}$ and $\tilde{\gamma}_{t}$. (An even simpler limiting form arises
in Theorem 1 in \S 3.1.) The relation involves the distribution of $Z_{s}^{%
\text{est}}:=\mathbb{E}[Z_{1}|Y_{s},\mathcal{G}_{t}^{+}],$ i.e of the time-$%
s $ estimator of the terminal output $Z_{1},$ conditional on the observation
of $Y_{s}$ by the agent. (In (\ref{LPM}) below, direct substitution causes a
spurious inner conditioning; the Landau notation below has the sense: $%
o(h)/h\rightarrow 0,$ as $h\downarrow 0.$)

\bigskip

\noindent \textbf{Proposition 1 (Conditional Bayes formula; }cf. \cite{JunK}%
\textbf{)}.\textit{\ In the single agent setting, conditional on there being
no disclosure in the interval }$(t,s),$ \textit{with }$\tilde{\gamma}_{t}:=%
\mathbb{E}[Z_{1}|\mathcal{G}_{t}^{+}]$ \textit{and }$\tilde{\gamma}_{s}:=%
\mathbb{E}[Z_{1}|Y_{s}=\gamma _{s},\mathcal{G}_{t}^{+}]$\textit{, the
equation}%
\begin{equation}
\tilde{\gamma}_{s}=\mathbb{E}[Z_{1}|ND_{s}(\gamma _{s}),\text{ }\mathcal{G}%
_{t}^{+}]  \label{Dye-1}
\end{equation}%
\textit{is equivalent for }$q_{ts}:=(s-t)\lambda _{t}$\textit{\ to}%
\begin{eqnarray}
(1-q_{ts})(\tilde{\gamma}_{t}-\tilde{\gamma}_{s})+o(s-t)
&=&q_{ts}\int_{z\leq \tilde{\gamma}_{s}}(z-\tilde{\gamma}_{s})\mathrm{d}%
\mathbb{Q}(Z_{s}^{\text{est}}\leq z|\mathcal{G}_{t}^{+})  \nonumber \\
&=&q_{ts}\int_{z\leq \tilde{\gamma}_{s}}(z-\tilde{\gamma}_{s})\mathrm{d}%
\mathbb{Q}(\mathbb{E}[Z_{1}|Y_{s},\mathcal{G}_{t}^{+}]\leq z|\mathcal{G}%
_{t}^{+}).  \label{LPM}
\end{eqnarray}

\bigskip

The proof is in \textbf{\S }5.1. The `indifference choice' of $\gamma $ is
of significance: we cite our earlier result here as:

\bigskip

\noindent \textbf{Proposition 2 (Risk neutrality, }\cite{OstG} -- cf. \cite%
{GieOb}\textbf{)}.\textit{\ In the setting of Proposition 1, with }$%
D=D_{s}(\gamma )$ \textit{and }$\tau _{D}^{t}:=\mathbb{Q}[D|\mathcal{G}%
_{t}^{+}]$\textit{, its market probability (conditional at time }$t$\textit{%
), equation }(\ref{Dye-1})\textit{\ is equivalent to }

\begin{equation}
S_{t}=\mathbb{E}[S_{1}|\mathcal{G}_{t}^{+}]=\tau _{D}^{t}\cdot \mathbb{E[}%
S_{1}|D_{s}(\gamma _{s}),\mathcal{G}_{t}^{+}]+(1-\tau _{D}^{t})\tilde{\gamma}%
_{s}.  \label{risk-neutral}
\end{equation}

Consequently, the computation of the (unique) solution for $\tilde{\gamma}%
_{s}$ reduces in the Black-Scholes setting to a simple application of the
Black-Scholes formulas, using the model parameters of \S 2.1; see the
discussion of Theorem 1 and the equivalent equation (\ref{Dye}) below in \S %
3.1.1.

\textbf{Remark.} From a market-valuation perspective on asset prices, the
value of the future cutoff $\gamma _{s}$ (as above) must be impounded in the
market measure, but this is exactly what (\ref{risk-neutral}) describes; so
we may validly regard the risk-neutral measure here as a summary of an
underlying equilibrium market-model, such as is described by \cite{DanJ}.

\subsection{Informal examples of censoring filters}

In the simplest context of $m=1$, take $\alpha _{1}=1$ and $f_{1}=1,$ so
that $Z_{1}=X_{1}.$ Given a filtration $\mathbb{G}$ generated via a
censoring filter $\gamma _{t}$ as above, we may term observations of $Y_{t}$
relative to $\gamma _{t}$ as \textit{bad} news at time $t$ when below $%
\gamma _{t},$ and as \textit{good} news at time $t$ when above or at $\gamma
_{t}.$ Omitting mention of the time, we refer to good and bad news relative
to $\tilde{\gamma}_{0},$ where $\tilde{\gamma}_{0}:=\tilde{m}_{0}(Y_{0})=%
\mathbb{E}_{0}[Z_{1}|Y_{0}]$ is the ex-ante valuation of the output process,
given the initial mandatory disclosure of $Y_{0}$ at time $t=0.$ We use the
notation%
\[
\tilde{m}_{t}(y)=\mathbb{E}_{t}[Z_{1}|Y_{t}=y]=\mathbb{E}_{t}[Z_{1}|Y_{t}=y,%
\mathcal{G}_{t}] 
\]%
for the relevant regression function, conditional on a time-$t$ disclosure $%
y;$ here the available information from disclosures dated before time $t$ is
indicated by the informal subscript in the expectation operator, formalized
as in \S 2.2 as conditioning on $\mathcal{G}_{t}$, so that%
\[
\mathbb{E}_{t}[.|...]:=\mathbb{E}[.|...,\mathcal{G}_{t}]. 
\]%
One presumes that $\tilde{m}_{t}(.)$ incorporates the \textit{improving}
precision over time $t$ of the observation process. In this context one may
discuss possible forms of suppression of bad news (relative to $\tilde{\gamma%
}_{0}),$ starting with two polar extremes: disclosure of all bad news (no
suppression) versus suppression of all bad news. Neither of these can be
supported by equilibrium considerations, but a natural third candidate,
incorporating a suitable, downward, time-varying risk-premium adjustment
format, is capable of equilibrium support -- see Theorem 1 below.

\bigskip

\noindent \textit{Example (Suppression-risk adjustment).}\textbf{\ }An
economically justified piecewise-deterministic approach to censoring (for
the case $m=1$) modifies the idea of naive `below the mean' suppression, by
anticipating how investors factor into their risk-neutral valuation a 
\textit{suppression-risk premium}. This leads to a downward adjustment of
the censor.

Two complementary effects underlie this premium-factor approach. To
understand this, suppose that the first disclosed observation occurs at the
stopping-time $\theta _{1}:=\inf \{t>0:$ $\tilde{m}_{t}(Y_{t}^{\text{obs}%
})\geq \tilde{\gamma}_{0}\}.$ Subsequently, until a further disclosure
occurs, a time-dependent downgrade factor should be applied for $t>\theta
_{1}$ to the mean $\tilde{m}_{\theta _{1}}=\tilde{m}_{\theta _{1}}(Y_{\theta
_{1}}^{\text{obs}})$ in recognition of two features: firstly, the
possibility, as time evolves, of undisclosed observations occurring below
the mean $\tilde{m}_{\theta _{1}}$ (with consequent lower conditional
expected output valuation), and secondly, the improved accuracy in the
forecasts of the output value $Z_{1}$ (by virtue of being closer to terminal
time). Consequently, the observation cutoff $\gamma _{t}$ satisfies $\tilde{m%
}_{t}(\gamma _{t})<\tilde{m}_{\theta _{1}},$ for $t>\theta _{1},$ which --
paradoxically -- implies that observations leading to output valuations
below the conditional mean may, after all, be disclosed; however, these give
rise to a higher valuation than would otherwise arise from the downgraded
mean. A secondary consideration is encouragement, as time evolves, for a
valuation-maximizing agent to make a fresh disclosure; this effect must
recognize that at any time $t$ there is a minimal expected valuation of the
terminal output, conditional on the uncertainty as to whether an observation
was suppressed.

The risk-premium calculation is driven by the determinism of the Poisson
arrival rate $\lambda _{t},$ and the multiplicative nature of the three
processes: output, observation, and state. In effect the risk-premium turns
out to be a deterministic zero-coupon bond associated with a risky asset,
valued at $\mathbb{E}_{t}[Z_{1}|Y_{t}]$; the bond here is a price function $%
B $ of two variables, $B=B(t,s),$ defined for $0\leq t\leq s<1,$ such that $%
B(t,t)=1,$ and $B(t,u)B(u,s)=B(t,s)$ for $0\leq t\leq u\leq s<1.$

In consequence, introducing `disclosure times' by reference to $\gamma _{t}$%
, inductively starting with $\theta _{1}:=\inf \{t>0:$ $N(t)>N(t-)$ \& $%
Y_{t}^{\text{obs}}\geq \gamma _{t}\}$, the `risky asset' is priced at $%
\tilde{\gamma}_{t}=\tilde{\gamma}_{\theta _{-}(t)}\cdot B(\theta _{-}(t),t)$%
; here as earlier $\theta _{-}(t)$ denotes the last disclosure time at or
below $t$. That is, $\tilde{\gamma}_{t}$ arises as though through a
bond-like `forward pricing' mechanism. Here $\tilde{\gamma}_{t}=\tilde{m}%
_{t}(\gamma _{t}),$ with $\gamma _{t}$ the cutoff for disclosure of $Y_{t}$.
Reference to consecutive down-crossing times, using such a formula, readily
yields an inductive construction of $\mathbb{G}$ as a subfiltration of $%
\mathbb{Y}^{\text{priv}}$.

The lesson of this example is two-fold. Firstly, the censoring filter $%
\gamma _{t}$ for $Y_{t}$ is a piecewise-deterministic Markov process in the
sense of M. H. Davis \cite{Dav2}, because $\tilde{m}_{t}(.)\ $is a
deterministic function. Secondly, one may readily describe the `disclosure'
filtration determined by $\gamma _{t}$ from $\mathbb{Y}^{\text{priv}}$ by an
inductive construction, as a subfiltration of $\mathbb{Y}^{\text{priv}}$
generated by the stopping-times defined above, $\theta _{1},\theta _{2},...$%
. Indeed, this points towards an alternative formalization of public
filtrations from stopping-times.

\section{Intra-period valuation: non-disclosure decay}

In this section\textbf{\ }a consistently generated filtration $\mathbb{\bar{G%
}}$ with its censoring filter $\gamma $ as defined in \S 2.1.7 is given and
assumed to be an \textit{optimal} censoring filter in the sense of (OC). The
latter is seen to be uniquely characterized as a piecewise-deterministic
Markov process in the sense of \cite{Dav2} (inevitably so -- see \cite{CalN}
and \cite{KurN}) capable of generating a public filtration from $\{\mathbb{Y}%
_{i}^{\text{priv}}:i\in I\}$ under which the censoring process is
predictable. First, we consider the simpler case $m=1$ (leaving the general
case $m>1$ to \S 3.2), and state the theorem asserting an \textit{explicit}
solution to the filtering problem (arising from a `cutoff equation'
characterizing $\gamma _{s}$, for $t<s<\theta _{+},$ which takes the form of
a simple differential equation). The proof of the theorem is in \textbf{\S }%
5.2, but we comment after the statement that the optimal censor satisfies a
Bayesian updating rule at time $s$, from which our differential equation
follows.

\subsection{Single agent case}

The situation when $m=1,$ and only the $i$-th agent is involved, follows. As
in \S 2.3, between public disclosure dates, we must refer not only to the
censor $\gamma _{t}$ which is applied to the observation $Y_{t}$, but also
to the image process%
\[
\tilde{\gamma}_{t}:=\mathbb{E}[Z_{1}|Y_{t}=\gamma _{t},\mathcal{G}_{t}]=%
\mathbb{E}[Z_{1}|ND_{t}(\gamma _{t}),\mathcal{G}_{t}]. 
\]%
In the Black-Scholes framework the time-$t$ regression function $\gamma
\mapsto \mathbb{E}[Z_{1}|Y_{t}=\gamma ,\mathcal{G}_{t}]$ is monotonic. In
the \textit{single} agent context, since the single observer is the only
source of any expansion of the public filtration, it is in principle
possible to work in the language of disclosure/censoring solely of the
forecast $\mathbb{E}[Z_{1}|Y_{t},\mathcal{G}_{t}]$ by reference to $\tilde{%
\gamma}_{t}$, rather than of the disclosure/censoring of the observation $%
Y_{t}$. However, in the multiple agent setting this cannot readily be done,
since other agents may expand the public filtration and so the connection
(in equilibrium) between $\tilde{\gamma}_{t}^{i}$ and $\gamma _{t}^{i}$ is
more complicated. So one simply has to chase both sets of variables. For
technical reasons connected with `re-starting' the Wiener process (at $t$
and at $\theta _{-}^{\text{pub}}(t))$, the censors need to be re-scaled to
unity at the re-starting date, hence the appearance also of a further
process $\hat{\gamma}_{t}$ in the Corollary below. (See \S 2.1.4 for the
relevant parameters.) In Theorem 1 reference is made to a fixed public
filtration $\mathbb{\bar{G}}$ and also to general public filtrations $%
\mathbb{G}$ as in (OC) in \S 2.2 above; moreover, a connection is made
between the \textit{optional valuation} $\mathbb{E}[Z_{1}|\mathcal{\bar{G}}%
_{t}^{+}]$ at the disclosure in (i), and the predictable valuation process $%
\mathbb{E}[Z_{1}|\mathcal{\bar{G}}_{t}]$ in (ii) -- cf. \S 2.2. Below $\Phi $
denotes the standard normal distribution.

\bigskip

\textbf{Theorem 1 (Decay Rule for }$m=1;$\textit{\ with only the} $i$-%
\textit{th agent present}\textbf{).} \textit{For the model of \textbf{\S }2,
suppose that }$\gamma _{t}$\textit{\ is a c\`{a}dl\`{a}g optimal
observation-censoring filter, generating a disclosure filtration }$\mathbb{%
\bar{G}}=\{\mathcal{\bar{G}}_{t}^{+}\}$ \textit{with associated sequence of }%
$\mathbb{\bar{G}}$\textit{-disclosure arrival times\ }$0=\theta _{0}^{\text{%
pub}}<\theta _{1}^{\text{pub}}<\theta _{2}^{\text{pub}}<...<\theta _{\ell }^{%
\text{pub}}<1,$\textit{\ of random (finite) length }$\ell \leq N(1)$ \textit{%
at which disclosures occur.}

\textit{The corresponding output-forecast process has the following
properties:}

\noindent (i) \textit{disclosure updating condition (`re-initialization')}%
\begin{equation}
g_{\ast }(\theta _{-}^{\text{pub}}(t)):=\mathbb{E}[Z_{1}|\mathcal{\bar{G}}%
_{t}^{+}]\equiv \mathbb{E}[Z_{1}|\mathcal{\bar{G}}_{t},Y_{t}],\text{ if }%
t=\theta _{-}^{\text{pub}}(t),  \label{re-initialize}
\end{equation}%
\textit{\ or explicitly here, for }$t=\theta _{-}^{\text{pub}}(t),$ 
\[
\mathbb{E}[Z_{1}|\mathcal{\bar{G}}_{t}^{+}]=\mathbb{E}[Z_{1}|Y_{t},\mathcal{%
\bar{G}}_{t}]=kY_{t}^{\kappa }, 
\]%
\textit{with }%
\[
k=f^{1-\kappa },\text{ }\kappa =\kappa _{1}^{i}=p_{i}/(p_{i}+p_{0}); 
\]%
\noindent (ii)\textit{\ in each inter-arrival interval (i.e. between
disclosure times)}%
\[
\sup\nolimits_{\mathbb{G}}\mathbb{E}[Z_{1}|\mathcal{G}_{t}]=\mathbb{E}[Z_{1}|%
\mathcal{\bar{G}}_{t}]=g_{\ast }((\theta _{-}^{\text{pub}}(t))\exp \left(
-\int_{\theta _{\_}}^{t}\nu _{s}ds\right) ,\text{ for }\theta _{-}^{\text{pub%
}}(t)\leq t<\theta _{+}^{\text{pub}}(t), 
\]%
\textit{where, on the extreme left of the display above, the
left-sided-in-time supremum} (\textit{as in }(OC))\textit{\ ranges over all
public filtrations }$\mathbb{G}=\{\mathcal{G}_{t}^{+}\}$\textit{\ and, on
the extreme right: }$g_{\ast }$ \textit{is as in }(\textit{i}) \textit{%
above, while in the valuation formula there is a thinned decay-intensity
given by}%
\begin{equation}
\nu _{t}=\lambda _{t}[2\Phi (\hat{\sigma}_{t}/2)-1]>0\text{ with }\hat{\sigma%
}_{t}^{2}=\alpha _{i}^{2}(\sigma _{0}^{2}+\sigma _{i}^{2})(1-t);
\label{mu-t}
\end{equation}%
\noindent (iii)\textit{\ in each inter-arrival interval, the cutoff }$\gamma
_{t}$\textit{\ for the disclosure of an observation of }$Y_{t}$\textit{\
satisfies}%
\[
\sup\nolimits_{\mathbb{G}}\mathbb{E}[Z_{1}|\mathcal{G}_{t}]=k\beta
_{t}\gamma _{t}{}^{\kappa }, 
\]%
\textit{where }$\beta =\tilde{\beta}^{i}:[0,1]\rightarrow \mathbb{R}$\textit{%
\ }(\textit{with }$\beta (1)=1)$ \textit{is a decreasing deterministic
weighting function of time, identified explicitly in Lemma 2 of \textbf{\S }%
5.2.}

\textit{In particular, the optimal censor }$\gamma _{t}$ \textit{is a
unique, piecewise-deterministic Markov process.}

\bigskip

The proof is in \textbf{\S }5.2. The cutoff for disclosure of the
output-forecast obtained in Theorem 1 has the prescribed explicit form,
since $\tilde{\gamma}_{t}=\sup\nolimits_{\mathbb{G}}\mathbb{E}[Z_{1}|%
\mathcal{G}_{t}]$ satisfies the \textit{censoring differential equation} 
\begin{equation}
\tilde{\gamma}_{t}^{\prime }=-\tilde{\gamma}_{t}\nu _{t},\text{ for }\theta
_{n-1}^{\text{pub}}<t<\theta _{n}^{\text{pub}},  \label{basic-de}
\end{equation}%
referred to in \S 1. It may be interpreted in the context of \S 2.3 as
expressing the \textit{risk premium} of information suppression, in a way
which hints at generalization to a broader class of models, one where regime
shifts are accompanied by optional, so strategic, `protective' activity,
their exercise rates balancing the marginal protective-option value. The
following result explains how `silence' (non-disclosure) is penalized less
and less as time plays out.

\bigskip

\textbf{Corollary. }\textit{In the setting of Theorem 1, the output-forecast
process }$\tilde{\gamma}_{t}$ \textit{for }$0<t<1$ \textit{has the following
properties:}

(a)\textit{\ its jumps occur at the disclosure times }$\theta _{n}^{\text{pub%
}}$ \textit{and are upward;}

(b)\textit{\ between jumps, the output-valuation process has the
representation }$\tilde{\gamma}_{t}=\hat{\gamma}_{t}g_{\ast }(\theta _{n-1}^{%
\text{pub}}),$\textit{\ where the (rescaled cutoff) deterministic function }$%
\hat{\gamma}_{t}$ \textit{satisfies:}%
\[
\hat{\gamma}_{t}^{\prime }=-\hat{\gamma}_{t}\nu _{t},\text{ for }\theta
_{n-1}^{\text{pub}}<t<\theta _{n}^{\text{pub}},\text{ with }\hat{\gamma}%
(\theta _{n-1}^{\text{pub}})=1; 
\]

(c) \textit{(decreasing thinning) in any interval between disclosure times
the relative decay-intensity }$\nu _{t}/\lambda _{t}$\textit{\ is decreasing;%
}

(d)\textit{\ between consecutive non-disclosure intervals the intra-period
relative decay intensity }$\nu _{t}/\lambda _{t}$ \textit{decreases (to zero
as }$t\rightarrow 1$\textit{).}

\bigskip

\noindent \textbf{Proof. }This is immediate -- the routine proof is omitted. 
$\square $

\subsubsection{Game-theoretic Aspects of Theorem 1}

We stress the role of game-theoretic principles underlying the proof: the
indifference principle, Bayesian updating, equilibrium, . Given information
at time $0<t<1,$ the cutoff value $\gamma _{s}$ is characterized at any time 
$s$ with $t<s<\theta _{+}(t)\leq 1$ by the observer's indifference at time $%
s,$ when observing $Y_{s},$ between disclosing the observed value if $%
Y_{s}=\gamma _{s}$ and not disclosing it; this is because the public
valuation is identical in both circumstances. Indeed, the valuation is
identical, because in the time interval $(t,s]$ with probability $%
1-(s-t)\lambda _{t}+o(s-t)$ (as $s\downarrow t$) the agent has not observed $%
Y_{s}$, and this event cannot be distinguished by outsiders from the event
that the agent observed $Y_{s}$, but did not disclose the observed value of $%
Y_{s}$ (since it was below $\gamma _{s}$). This yields the indifference
(equilibrium) condition as a \textit{conditional Bayes formula: } 
\begin{equation}
\tilde{\gamma}_{s}=\mathbb{E}[Z_{1}|ND_{s}(\gamma _{s}),\mathcal{G}_{t}^{+}],
\label{Dye}
\end{equation}%
where, as earlier $\tilde{\gamma}_{s}:=\mathbb{E}[Z_{1}|Y_{s}=\gamma _{s},%
\mathcal{G}_{t}^{+}]$ and $\mathbb{G}=\{\mathcal{G}_{t}^{+}\}$ here denotes
the public filtration.

The significance of (\ref{Dye}) is that it is a special case of the \textit{%
Nash Equilibrium} condition (\ref{NE}) below; the equation (\ref{Dye}) first
appears in the static model of Dye \cite{Dye} to model rationality of
partial (voluntary) disclosures, as a contrast to the total disclosure
principle (`unravelling') of Grossman and Hart \cite{GroH}.

\subsubsection{Cutoffs and the Black-Scholes formula}

A consequence of (\ref{Dye}) (and of (\ref{NE}) below) and is a
Black-Scholes formula for the cutoff $\gamma =\tilde{\gamma}_{s}$ applied to
the forecast (rather than the observation). The calculation goes back at
least to \cite{JunK} (cf. \cite{GieOa}), where (\ref{Dye}) in the present
context reduces to:%
\begin{eqnarray}
\mu _{F}-\gamma &=&\frac{q}{1-q}H_{F}(\gamma ),\text{ where}
\label{JungKwon7} \\
H_{F}(\gamma ) &=&\mathbb{E}[(\gamma -F)^{+}]=\int (\gamma -x)^{+}d\mathbb{Q(%
}F\leq x)=\int_{x\leq \gamma }\mathbb{Q(}F\leq x)dx,  \nonumber
\end{eqnarray}%
(cf. Prop. 1 and \textbf{\S }5.1). Here $F$ denotes the random variable $%
\mathbb{E}[Z_{1}|Y_{s},\mathcal{G}_{t}^{+}]$, with mean $\mu _{F}$
(conditional at time $t+),$ $q=(s-t)\lambda _{t}$ is the probability with
which an observation occurs (independently of $F$) by time $s$, and $%
H_{F}(\gamma )$ is the `lower first partial moment below a target $\gamma $%
', well-known in risk management\footnote{%
See for example McNeil, Frey and Embrechts \cite{McNFE}, Section 2.2.4.},
briefly termed (in view of its key role) the \textit{hemi-mean function}.

The log-normal $F$ above, prompts a standardization in terms of the
parameters $\lambda ,\sigma $ for the solution $\gamma =\gamma _{\text{LN}%
}(\lambda ,\sigma )$ of%
\begin{equation}
1-\gamma =\lambda H_{\text{LN}}(\gamma ;\sigma ).  \label{H-LN}
\end{equation}%
The behaviour of the solution derives from the properties of the
Black-Scholes \textit{put} formula as a function of its strike --
equivalently of the corresponding \textit{call} formula as a function of its
underlying asset price (Black-Scholes `put-call symmetry', cf. [BarNS, \S\ %
11.4], [BjeS], [CarL] -- for background see [Teh]). For strike $\gamma $ and
expiry at time $1,$ conditional on an initial asset valuation of $\tilde{%
\gamma}_{t}$ at time $t<1,$ the formula reads

\[
H_{\text{LN}}(\gamma )=\gamma \Phi \left( \frac{\log (\gamma /\tilde{\gamma}%
_{t})+\frac{1}{2}\sigma ^{2}(1-t)}{\sigma \sqrt{1-t}}\right) -\gamma
_{t}\Phi \left( \frac{\log (\gamma /\tilde{\gamma}_{t})-\frac{1}{2}\sigma
^{2}(1-t)}{\sigma \sqrt{1-t}}\right) , 
\]%
with $\Phi $ the standard normal distribution as above.

Consequently, at any time $t$ between consecutive voluntary disclosures,
setting the strike above to $\gamma =\tilde{\gamma}_{s}$ yields in the limit
as $s\downarrow t$ (with $\tilde{\gamma}_{s}\rightarrow \tilde{\gamma}_{t}$)
an \textit{at-the-money }limiting\textit{\ forward-start call} formula; this
at-the-money option-value simplifies to $\tilde{\gamma}_{t}2\Phi (\sigma
_{t}/2)$, as in (\ref{mu-t}). We hope to return to this aspect elsewhere.

\subsubsection{Other Comments}

The decay-intensity rate $\nu _{t}$ is composed of two factors both having
economic significance. First, the Poisson intensity $\lambda $ measures the
instantaneous opportunity cost of the arrival of an observation of
information about the output valuation, and impounds the chance both for a
valuation upgrade (through a good observation), and for the suppression of
poor observation value. Secondly, the intensity $\lambda $ is thinned by the
probability of suppressing poor valuation, the effect of a `protective put'.
Over time the protective put loses value and tends to zero as the precision
improves (i.e. the volatility goes to zero).

To see the details of the formula intuitively, recall that, as above, it is
assumed that the observer chooses to achieve the maximum valuation, and thus
makes a voluntary disclosure according to the rule: find the cutoff function 
$t\mapsto \gamma _{t}$ such that for $t=\theta _{n}^{\text{obs}}:$

\noindent (i) disclose credibly the value of $Y$, when $Y_{t}\geq \gamma
_{t},$ or, equivalently, the public output-forecast $Z_{t}^{\text{est}}:=%
\mathbb{E}_{t}[Z_{1}|Y(\theta _{n}^{\text{obs}})];$ and

\noindent (ii) make no disclosure, when $Y_{t}<\gamma _{t}$.

\noindent Replacing $Y_{t}$ by $Z_{t}^{\text{est}}:=\mathbb{E}[Z_{1}|%
\mathcal{G}_{t}^{+}],$ one may restate the $Y$-cutoff problem (of choosing $%
\gamma _{t}$) in isomorphic terms as a cutoff problem for $Z_{t}^{\text{est}%
} $ (i.e. of choosing its corresponding cutoff $\tilde{\gamma}_{t})$,
assuming the regression function $\tilde{m}_{t}(y):=\mathbb{E}%
_{t}[Z_{1}|Y_{t}=y]$ is monotonic.

\subsection{Multiple agent case}

We now consider the general case $m>1$. Here the starting point is a system
of equations that relates the values $\{\gamma _{t}^{i}:i=1,...,m\}$ to each
other by reference to a general Bayesian updating rule having the form of a
system of (subgame) \textit{Nash equilibrium} conditions for $i=1,...,m$:%
\begin{equation}
\mathbb{E}[Z_{1}^{i}|(\forall j)ND_{t}^{j}(\gamma _{t}^{j}),\mathcal{G}_{t}]=%
\mathbb{E}[Z_{1}^{i}|(\forall j\neq i)ND_{t}^{j}(\gamma
_{t}^{j})ND_{t}^{j}(\gamma _{t}^{j}),Y_{i}=\gamma _{t}^{i},\mathcal{G}_{t}],
\label{NE}
\end{equation}%
with $ND_{t}^{j}(\gamma )$ being the non-disclosure event (of time $t$). The
intended meaning is that, contingent on the information available prior to
time $t,$ in the event that all of the agents make no disclosures (for lack
of observations, or because observations lie at or below their respective
filtering-censor value) and the $i$-th agent's observation is identical to
the value of the respective filtering-censor $\gamma _{t}^{i}$, the
corresponding output estimate value (i.e. $Z_{1}^{i}$ in expectation) is the
same whether, or not, that agent chooses to disclose the observation.
(Recall the equational convention in the footnote of \S 2.3.) The conditions
(\ref{NE}) generalize (\ref{risk-neutral}) of \S 2.3 -- see also \S 3.1.1
above.

Note that $ND_{t}^{j}(\gamma )$ is complementary to $D_{t}(\gamma ),$ as
earlier defined in \S 2.3 (though here in respect of the $j$-th agent).

In the inter-arrival period the absence of any disclosure from all of the $m$
observation processes will influence the decay rate of each of the optimal
observation filtering censors differentially, i.e. the filtering equation is
bound to express the interdependence flowing from the Nash Equilibrium
conditions (\ref{NE}) above. To express the explicit form, we need a number
of parameters derived from the volatilities of \S 2.1. Using an abbreviating 
\textit{tilde notation}, put%
\[
\sigma _{0i}^{2}:=\sigma _{0}^{2}+\sigma _{i}^{2}\text{ and }\tilde{\sigma}%
_{0i}^{2}(t):=(1-t)\sigma _{0i}^{2}, 
\]%
and analogously: 
\begin{eqnarray*}
\tilde{p}_{i}(t) &:&=1/[(1-t)\sigma _{i}^{2}],\text{ }\tilde{p}%
(t):=\sum\nolimits_{i=0}^{m}\tilde{p}_{i}(t),\quad \tilde{\kappa}_{i}(t):=%
\tilde{p}_{i}(t)/\tilde{p}(t)\equiv \kappa _{i},\text{ } \\
\text{ }\tilde{\kappa}_{-i}(t) &:&=\tilde{p}_{i}(t)/(\tilde{p}(t)-\tilde{p}%
_{i}(t))\equiv \kappa _{-i}:=p_{i}/(p-p_{i}).
\end{eqnarray*}%
Of particular significance is the function $\tilde{\rho}_{i}(t),$ which
denotes for given $t$ the (conditional) partial covariance (cf. \cite{CoxS})
of the $i$-th component of $(...,\sigma _{0}\tilde{W}_{1-t}^{0}+\sigma _{i}%
\tilde{W}_{1-t}^{i},...)$ on the remaining components, where $\tilde{W}$
denotes the Wiener process $W$ re-started at time $t.$

We may now state the general theorem; use of the subscript `hyp' here is
explained in the discussion below. The function $\tilde{\beta}_{m}^{i}$
corresponds to $\tilde{\beta}^{i}$ in Theorem 1, and is derived in Lemma 2$%
_{m}$ of \textbf{\S }5.4. Below $\tilde{\gamma}_{it}=\sup\nolimits_{\mathbb{G%
}}\mathbb{E}[Z_{1}^{i}|\mathcal{G}_{t}],$ is as earlier, but $y_{it}$
replaces $\gamma _{it}$ to allow $\tilde{y}_{it}$ to have another meaning,
corresponding to a rescaling of $Y_{s}^{i}$ to $\tilde{Y}_{s}^{i}$ in the
proof. The notation here of $\mathbb{\bar{G}}$ and $\mathbb{G}$ is similar
to that in Theorem 1.

\bigskip

\textbf{Theorem 1}$_{m}$\textbf{\ (Filtering rule during continued
non-disclosure: }$m\geq 1$\textbf{).} \textit{For the model of \textbf{\S }%
2, suppose }$y_{t}=(...,y_{it},...)$\textit{\ is a c\`{a}dl\`{a}g optimal
censoring filter generating a disclosure filtration }$\mathbb{\bar{G}}$ 
\textit{with associated sequence of }$\mathbb{\bar{G}}$\textit{%
-disclosure-arrival times \ }$0=\theta _{0}^{\text{pub}}<\theta _{1}^{\text{%
pub}}<\theta _{2}^{\text{pub}}<...<\theta _{\ell }^{\text{pub}}<1$\textit{\
of random (finite) length }$\ell \leq \sum_{i=1}^{m}N^{i}(1),$ \textit{at
which disclosure events occur.}

\textit{Then, for }$0\leq n\leq \ell ,$ \textit{the corresponding
output-forecast process has the following properties:}

\noindent (i) \textit{disclosure updating condition (the }$i$\textit{-th
agent's `re-initialization') }%
\begin{equation}
g_{\ast }^{i}(t):=\mathbb{E}[Z_{1}^{i}|\mathcal{G}_{t}^{+}]=\mathbb{E}%
[Z_{1}^{i}|\{Y_{t}^{j}:j\in J_{t}\}],\text{ if }t=\theta _{-}^{\text{pub}%
}(t);  \label{re-initial}
\end{equation}%
\noindent (ii) \textit{in each inter-arrival interval }$\theta =\theta _{-}^{%
\text{pub}}(t)\leq t<\theta _{+}^{\text{pub}}(t)$\textit{\ }%
\[
\sup\nolimits_{\mathbb{G}}\mathbb{E}[Z_{1}^{i}|\mathcal{G}_{t}]=\mathbb{E}%
[Z_{1}^{i}|\mathcal{\bar{G}}_{t}]=k_{m}^{i}\tilde{\beta}_{m}^{i}g_{\ast
}^{i}(\theta )\exp \left( -\int_{\theta }^{t}\nu _{\text{agg}}(s)ds\right) , 
\]%
\textit{where on the extreme left, the left-sided-in-time supremum ranges
over public filtrations }$\mathbb{G}=\{\mathcal{G}_{t}^{+}\}$ \textit{and,\
on the extreme-right, the correlation-aggregated decay-intensity }$v_{\text{%
agg}}$\textit{\ is}%
\[
\nu _{\text{agg}}(t):=\sum\nolimits_{j}\frac{\kappa _{j}}{\kappa _{-j}}%
\left( 1+\sum\nolimits_{h}\frac{\alpha _{h}}{\alpha _{j}}\frac{\kappa _{h}}{%
\kappa _{0}}\right) \nu _{j\text{hyp}}(t), 
\]%
\textit{and}%
\[
\nu _{i\text{hyp}}(t):=\Phi \left( \frac{1}{2}\alpha _{i}\kappa _{i}\tilde{%
\sigma}_{0i}\sqrt{1-\tilde{\rho}_{i}^{2}}\right) \tilde{\lambda}_{i}, 
\]%
\textit{and }$\tilde{\beta}_{m}^{i}=\tilde{\beta}_{\text{indiv}}^{i}\cdot 
\tilde{\beta}_{\text{agg}}$ \textit{with}%
\[
\tilde{\beta}_{\text{indiv}}^{i}:=\mu (\alpha _{i},(1-\kappa _{0})\tilde{%
\sigma}_{0}^{2})\mu (\kappa _{1}^{i},(1-\kappa _{0})\tilde{\sigma}_{i}^{2}),%
\text{ and }\tilde{\beta}_{\text{agg}}=\prod\nolimits_{j}\mu (\kappa _{j},%
\tilde{\sigma}_{0j}^{2}); 
\]%
\noindent (iii) \textit{in each inter-arrival interval between disclosure
event times, the disclosure censor }$y_{t}^{i}$ \textit{of the }$i$\textit{%
-th agent is given for }$\theta =\theta _{n-1}^{\text{pub}}\leq t<\theta
_{n}^{\text{pub}}$ \textit{by:}%
\begin{eqnarray*}
\frac{1}{\alpha _{i}}\log y_{t}^{i} &=&-\frac{1}{\alpha _{i}\kappa _{-i}}%
\int_{\theta }^{t}\nu _{i\text{hyp}}(s)ds \\
&&-\frac{1}{\kappa _{0}}\left( \frac{\kappa _{1}}{\alpha _{1}\kappa _{-1}}%
\int_{\theta }^{t}\nu _{1\text{hyp}}(s)ds+\frac{\kappa _{2}}{\alpha
_{2}\kappa _{-2}}\int_{\theta }^{t}\nu _{2\text{hyp}}(s)ds+...\right) .
\end{eqnarray*}

\textit{In particular, the optimal censor }$y_{t}$ \textit{is a unique,
piecewise-deterministic Markov process.}

\bigskip

Note that for $m=1$ this reduces to Theorem 1. The proof is in \textbf{\S }%
5.3 and depends (as does Theorem 1) on the factorization of the process $%
M=M^{i}$ (with $\sigma _{M}$ for $\sigma _{M_{i}})$ for fixed $t$ in the
self-evident form%
\[
M_{1}=M_{0}\exp (\sigma _{M}W_{1}-\frac{1}{2}\sigma _{M}^{2})=M_{t}\exp
(\sigma _{M}\tilde{W}_{1-t}-\frac{1}{2}\sigma _{M}^{2}(1-t)), 
\]%
where $\tilde{W}^{i}$ is $W^{i}$ re-started at time $t.$

\bigskip

\noindent \textbf{Discussion of Theorem 1}$_{m}$\textbf{. }This result
builds on Theorem 1, hence bears appropriate similarities (e.g. updating at
disclosure dates), and relies on the conditional Bayes formula (\ref{Dye}),
so we now comment only on what is most significantly different here for $m>1$%
, namely the need to disaggregate the co-dependance. To define the
observation cutoffs of the $m$ observing agents, one first constructs $m$
corresponding agents, termed \textit{hypothetical} agents, each of whom
faces a suppressed-observation problem of the kind considered in Theorem 1,
but in isolation. This introduces two features: firstly, an \textit{%
amended-mean factor }$L_{-i}(t)$, multiplying the current conditional mean
of the observation process (defined in Theorem M\ of \textbf{\S }5.3 and
reflecting incremental effects of a single agent, that are based only on
loading and precision factors) -- and, secondly, the \textit{partial
covariance} $\tilde{\rho}_{i},$ arising from the use of the Schur complement
(see \cite[Note 4.27, p.120]{BinF}, cf. \cite[Ch. 27]{KenS-2}, \cite[\S \S %
46.26-28]{KenS-3}). The corresponding hypothetical observation-cutoffs are
later aggregated to yield observation cutoffs of the original $m$-agent
problem. From these observation cutoffs a current forecast $\tilde{\gamma}%
_{t}^{i}$ of the $i$-th output $Z_{1}^{i}$ is derived. Recall that in
Theorem 1 the formula for $\tilde{\gamma}_{t}$ required the use of a
rescaled function $\hat{\gamma}_{t}$ (with unit value at $t=\theta ,$ the
latest disclosure date). Likewise in Theorem 1$_{m}$ we also see a function $%
\hat{\gamma}_{t}^{i}$ used in similar fashion to construct $\tilde{\gamma}%
_{t}^{i};$ here the size-constant $k_{m}^{i}$ corresponds to $f^{i}$, and
reflects relative output magnitude, just as $\kappa _{i}$ reflects relative
precision.

\section{Application: Market valuation with censored voluntary disclosures}

The differential equations of Theorem 1$_{m}$ can be used to derive
comparative statics of price formation in an asset market, in which
investors expect voluntary disclosures (with positive probability) at all
times between the two fixed mandatory disclosure dates. These comparative
statics are concerned with intervals of time during which the agents are
known to be privately and intermittently observing noisy signals of the
asset values, arising from the common effect. The interpretation of \S 2.3
extends here to 
\[
S_{t}^{i}:=\mathbb{E}[\mathbb{E[}Z_{1}^{i}|\mathcal{G}_{1}^{+}]|\mathcal{G}%
_{t}^{+}] 
\]%
as an asset-price process (with expectations under $\mathbb{Q}$ as the
market's risk-neutral valuation measure).

Agent $i$ is endowed intermittently with private information about the
evolution of the $i$-th asset price via the martingale $Y_{t}^{i},$ and can
voluntarily disclose information about the asset to the market. At times $t$
strictly between consecutive public disclosure arrival times, conditioning
on $\mathcal{G}_{t}^{+}$ or on $\mathcal{G}_{t}$ yields identical forecasts
of the time-$1$ valuation $\mathbb{E[}Z_{1}^{i}|\mathcal{G}_{1}^{+}]$, so
correspondingly the asset price $S_{t}^{i}$ is identical with $\tilde{\gamma}%
_{t}^{i}=\mathbb{E}[\mathbb{E[}Z_{1}^{i}|\mathcal{G}_{1}^{+}]|\mathcal{G}%
_{t}],$ where $\tilde{\gamma}_{t}^{i}$ is the unique valuation-cutoff given
by Theorem 1$_{m}.$ Furthermore, \S 2.3 describes disclosure behaviour as
being incentivized so that the asset price gives the current stock-holders
(investors) asset valuations that are maximal given the available public
information.

The effects on price formation may thus be studied by recourse to the
observation cutoffs employed by the respective agents in terms of the
parameters of the model: the precisions $p_{i},$ the loading factors $\alpha
_{i}$, and the private observation arrival intensities $\lambda _{i}.$ It is
interesting to note the predictions about suppression. Since the correlation
between firms and the environment factor are positive, the formulas
established above imply that a good-news \textit{bandwagon} \textit{effect}
holds: ceteris paribus, agents \textit{all} choose a higher cutoff (relative
to the single agent case $m=1$),\ reducing the probability that they will
release private observations. Additionally, there is an intuitively clear 
\textit{estimator-quality effect,} which leads to agents being \textit{%
partitioned} into below- and above-`average precision'\textbf{\ }(over the $%
m $-agent population), as in the theorems that follow. Those with
below-average precision are shown to adopt a lower cutoff (relative to the
single agent case), and ceteris paribus increase the probability that they
will release private observations, with the reverse holding for the
above-average.

\bigskip

\noindent \textbf{Bandwagon Theorem. }\textit{In any intra-period the
presence of correlation increases the precision parameter of the cutoff and
hence raises the cutoff:} 
\[
\hat{\gamma}_{\text{LN}}(\tilde{\lambda}_{i},\tilde{\sigma}_{0i})<\hat{\gamma%
}_{\text{LN}}(\tilde{\lambda}_{i},\kappa _{i}\tilde{\sigma}_{0i})<\hat{\gamma%
}_{\text{LN}}\left( \tilde{\lambda}_{i},\kappa _{i}\tilde{\sigma}_{0i}\sqrt{%
1-\tilde{\rho}_{i}^{2}}\right) , 
\]%
\textit{where }$\hat{\gamma}_{\text{LN}}(\lambda ,\sigma )$\textit{\ denotes
the unique solution of the following equation in }$y:$%
\[
(1-y)=\lambda H_{\text{LN}}(y,\sigma ), 
\]%
\textit{\ and represents the normalized cutoff in the single agent case, as
in }(\ref{H-LN})\textit{.}

\bigskip

\noindent \textbf{Proof.} This is immediate from the static model of \cite[%
\S 6]{GieOa}.

\bigskip

When the correlation is positive, there is also a counter-vailing precision
effect on the related hypothetical agent's cutoff, arising from the amended
mean factor $L_{-i}$ (see the discussion of Theorem 1$_{m}$ above), when the
actual agent has below-average precision, as defined below.

\bigskip

\noindent \textbf{Estimator-Quality Theorem.} \textit{Suppose that }$m\geq 2$
\textit{and }$\alpha _{i}>0$\textit{\ for all }$i$. \textit{The amended mean 
}$L_{-i}(t)$\textit{\ of the hypothetical firm }$i$\textit{\ increases with }%
$p_{i}$\textit{\ and}%
\begin{eqnarray*}
\exp \left( -\frac{\alpha _{i}(1-t)}{2(p-p_{i})}\right) &<&L_{-i}(t)<\exp
\left( \frac{\alpha _{i}\left( 1+\frac{\alpha _{i}-1}{n-1}\right) (1-t)}{2p_{%
\text{av,}-i}}\right) ,\text{ } \\
\text{where }p_{\text{av,}-i} &:&=\frac{p-p_{i}}{n-1+\alpha _{i}}.
\end{eqnarray*}%
\textit{\ In particular, if the loading index is identical for all firms,
then}%
\[
L_{-i}(t)<L_{-j}(t)\text{ iff }p_{i}<p_{j}. 
\]%
\textit{Otherwise, if }$0<\alpha _{i}<\alpha _{j}$ \textit{and }$%
p_{i}<p_{j}, $\textit{\ then also }$L_{-i}(t)<L_{-j}(t)$\textit{.}

\textit{The amended mean is a strict deflator, i.e. }$L_{-i}(t)<1,$\textit{\
iff }$p_{i}$\textit{\ is below the loading-adjusted competitor average, i.e.}%
\[
p_{i}<\frac{p}{n-1+\alpha _{i}}:=p_{\text{av,-}i} 
\]%
\textit{so that for }$\alpha _{i}=1$\textit{\ one has }$p_{\text{av,-}%
i}=p/n. $

\bigskip

\noindent \textbf{Proof.} This again is immediate from the static model of 
\cite[\S 6]{GieOa}.

\bigskip

Thus low-precision managers are more likely to make a disclosure, but by
definition the disclosure will be less precise. Hence, in terms of giving it
weighting in the updating rules, investors will give less weight to
disclosure of bad news by such imprecise managers.

\section{Proofs}

We begin in \textbf{\S }5.1 with a Proof of Proposition 1 and use it in 
\textbf{\S }5.2 to prove Theorem 1, but only after we have prepared the
ground with the calculations in two lemmas. In \textbf{\S }5.3 we prove
Theorem 1$_{m};$ the argument will require generalizations of the lemmas of 
\textbf{\S }5.2, and these are relegated to \textbf{\S }5.4.

\subsection{Proof of Proposition 1}

\noindent In the notation of \textbf{\S }2, we have%
\begin{eqnarray*}
\tilde{\gamma}_{s} &=&\frac{\mathbb{Q}(N(s)=N(t))\cdot \mathbb{E}[Z_{1}|%
\mathcal{G}_{t}^{+}]+\mathbb{Q}(N(t)<N(s))\cdot \int_{z_{1}\leq \tilde{\gamma%
}_{s}}z_{1}d\mathbb{Q}(\mathbb{E}[Z_{1}|Y_{s},\mathcal{G}_{t}^{+}]\leq z_{1}|%
\mathcal{G}_{t}^{+})}{\mathbb{Q}(N(s)=N(t))+\mathbb{Q}(N(t)<N(s))\mathbb{Q}(%
\mathbb{E}[Z_{1}|Y_{s},\mathcal{G}_{t}^{+}]\leq \tilde{\gamma}_{s}|\mathcal{G%
}_{t}^{+})} \\
&=&\frac{(1-(s-t)\lambda _{t})\cdot \tilde{\gamma}_{t}+(s-t)\lambda
_{t}\cdot \int_{z_{1}\leq \tilde{\gamma}_{s}}z_{1}d\mathbb{Q}(\mathbb{E}%
[Z_{1}|Y_{s},\mathcal{G}_{t}^{+}]\leq z_{1}|\mathcal{G}_{t}^{+})}{%
(1-(s-t)\lambda _{t})+(s-t)\lambda _{t}\mathbb{Q}(\mathbb{E}[Z_{1}|Y_{s},%
\mathcal{G}_{t}^{+}]\leq \tilde{\gamma}_{s}|\mathcal{G}_{t}^{+})}.
\end{eqnarray*}%
Putting $q_{ts}=(s-t)\lambda _{t},$ expressing $\mathbb{Q}$ as $\int d%
\mathbb{Q}$, cross-multiplying and re-arranging%
\begin{eqnarray*}
(1-q_{ts})(\tilde{\gamma}_{t}-\tilde{\gamma}_{s})+o(s-t)
&=&q_{ts}\int_{z_{1}\leq \tilde{\gamma}_{s}}(z_{1}-\tilde{\gamma}_{s})%
\mathrm{d}\mathbb{Q}(\mathbb{E}[Z_{1}|Y_{s},\mathcal{G}_{t}^{+}]\leq z_{1}|%
\mathcal{G}_{t}^{+}) \\
&=&-q_{ts}\int_{z_{1}\leq \tilde{\gamma}_{s}}\mathbb{Q}(\mathbb{E}%
[Z_{1}|Y_{s},\mathcal{G}_{t}^{+}]\leq z_{1}|\mathcal{G}_{t}^{+})\mathrm{d}%
z_{1},
\end{eqnarray*}%
where the last line is from integrating by parts. $\square $

\subsection{Proof of Theorem 1}

This section is devoted to a proof of Theorem 1.

\bigskip

\textbf{Idea of the proof.} This rests on two observations. The first is
that, since the regression function $\tilde{m}_{t}(\gamma ):=\mathbb{E}%
[Z_{1}|Y_{t}=\gamma ,\mathcal{G}_{t}]$ is explicit (Lemma 1 below -- but
here we omit the agent's index) and monotone, the censoring behaviour
dictated by a cutoff function $\gamma _{t}$ for the disclosure of $Y_{t}$
may be equivalently expressed as a censoring function $\tilde{\gamma}_{t}$
for the disclosure of the valuation process $\tilde{Z}_{t}:=\tilde{m}%
_{t}(Y_{t}),$ where%
\[
\tilde{\gamma}_{t}=\tilde{m}_{t}(\gamma _{t}). 
\]%
The second -- the nub of the proof (from Prop. 1) -- is that $\tilde{\gamma}%
_{t}$ may be characterized by an equation expressing the agent's
indifference between non-disclosure and disclosure of $\tilde{Z}_{t}$ when $%
\tilde{Z}_{t}=\tilde{\gamma}_{t}.$ Such an equation gives $\tilde{\gamma}%
_{t} $ only \textit{implicitly}; however, a variational analysis of the
equation \textit{explicitly} determines $\tilde{\gamma}_{t}$ via an easy
differential equation. Inverting the regression yields $\gamma _{t}$ in
explicit form as%
\[
\gamma _{t}=\tilde{m}_{t}^{-1}(\tilde{\gamma}_{t}). 
\]%
In the variational analysis above, the agent treats $\tilde{Z}_{t}$ as a
noisy signal of $Z_{1}$ on the grounds that $\mathbb{E}_{t}[Z_{1}|\tilde{Z}%
_{t},\mathcal{G}_{t}]=\mathbb{E}_{t}[Z_{1}|Y_{t},\mathcal{G}_{t}]=\tilde{Z}%
_{t}$. This leads to a two-period analysis, in which the process value $%
Z_{1} $ is viewed, from the standpoint of information of time $t,$ as a
random variable, and permits exploitation of known results from two-period
analyses.

\bigskip

Although $m=1$ below, having in mind the notational needs of \textbf{\S }5.3$%
,$ where $m>1,$ it is helpful to employ here a general index $i$ for the
single agent of interest (rather than to specialize $i=1)$, thereby
anticipating the later general case. Also, as $m=1,$ the index $i$ can be
omitted at will, whenever convenient. We begin with some auxiliary results
stated as Lemmas 1 and 2, which will require the notation%
\[
\mu (\kappa ,\sigma ^{2}):=e^{\frac{1}{2}\kappa (\kappa -1)\sigma ^{2}}. 
\]%
The exponent $\frac{1}{2}\kappa (\kappa -1)\sigma ^{2}$ in $\mu $ above
corresponds to the second-order term of It\^{o}'s Lemma for the power
function $t\rightarrow t^{\kappa },$ a recurring feature of the regression
formulas (from Lemma 1). For other parameters refer to \S 2.1.

Recall the tilde notation of \S 3.2 that, for any constant $\sigma ,$ we
write $\tilde{\sigma}^{2}$ for the function $\tilde{\sigma}^{2}(t):=\sigma
^{2}\cdot (1-t),$ corresponding to re-starting the model at time $t.$ In
particular, with $\bar{\sigma}_{0i}^{2}:=\alpha _{i}^{2}\sigma
_{0}^{2}+(\alpha _{i}\sigma _{i})^{2}$, we write%
\[
\tilde{\sigma}_{i}^{2}(t)=\sigma _{i}^{2}(1-t)\text{ and }\bar{\sigma}%
_{0i}^{2}(t)=\alpha _{i}^{2}[\sigma _{0}^{2}+\sigma _{i}^{2}](1-t)=\alpha
_{i}^{2}\tilde{\sigma}_{0i}^{2}. 
\]%
For convenience, we may denote $X$ also by $M^{0}.$ For the Wiener processes 
$W^{i}$, recall the corresponding Wiener process re-started at time $t$%
\[
\tilde{W}_{s}^{i}:=W_{t+s}^{i}-W_{t}^{i}. 
\]%
(So $\tilde{W}_{0}=0$.) Omitting suffices and writing $\sigma _{W}$ for $%
\sigma _{i}^{M},$%
\begin{equation}
M_{t+s}=M_{t}\exp (\sigma _{W}\tilde{W}_{s}-\frac{1}{2}\sigma _{W}^{2}s).
\label{decomp}
\end{equation}

We need two lemmas.

\bigskip

\noindent \textbf{Lemma 1 (Valuation of }$Z_{1}$\textbf{\ given observation }%
$Y_{1}$ \textbf{for }$m=1$\textbf{).} \textit{Put}%
\[
\kappa =\kappa _{1}^{i}=p_{i}/(p_{0}+p_{i}). 
\]%
\textit{Then} 
\[
\mathbb{E}[Z_{1}|Y_{1}^{i}=y,\mathcal{G}_{1}]=ky^{\kappa }\text{ for }%
k=k_{1}^{i}=(f^{i})^{1-\kappa }. 
\]

\noindent \textbf{Proof. }We cite and apply a formula from \cite[Prop. 10.3,
with $n=1$]{GieOa}. To distinguish notational contexts, we use overbars on
letters when citing formulas from there. The noisy observation there, $\bar{T%
},$ of a random (state) variable $\bar{X}$ takes the form $\bar{T}=\bar{X}%
\bar{Y},$ where $\bar{X}$ and $\bar{Y}$ are independent random variables,
with their log-normal distributions having underlying normal precision
parameters $\bar{p}_{X},\bar{p}_{Y}$ respectively. The required formula is%
\[
\mathbb{E}[\bar{X}^{\alpha }|\bar{T}]=\bar{K}_{\alpha }(\bar{T})^{\alpha
\kappa }, 
\]%
where $\kappa =\bar{p}_{Y}/\bar{p}$ for $\bar{p}=\bar{p}_{X}+\bar{p}_{Y}$ and%
\[
\bar{K}_{\alpha }=\exp \left( \frac{\alpha +\alpha (\alpha -1)}{2\bar{p}}%
\right) . 
\]%
We will take $\bar{X}=X_{1}$ and $\bar{Y}=M_{1}^{1/\alpha },$ with $M=M^{i}$
and $\alpha =\alpha _{i}$ below. We first compute the corresponding
constants $\bar{p}_{X},\bar{p}_{Y}$ and $\bar{K}_{\alpha }.$ Substituting $%
s=1-t=\Delta t$ in (\ref{decomp}) above, gives%
\[
X_{1}=X_{1-\Delta t}\exp (\sigma _{0}\tilde{W}_{\Delta t}^{0}-\frac{\sigma
_{0}^{2}}{2}\Delta t), 
\]%
and%
\[
M_{1}^{1/\alpha }=M_{1-\Delta t}^{1/\alpha }\exp (\frac{\sigma _{i}^{M}}{%
\alpha _{i}}\tilde{W}_{\Delta t}^{i}-\frac{(\sigma _{i}^{M})^{2}}{2\alpha
_{i}}\Delta t)=M_{1-\Delta t}^{1/\alpha }\exp (\sigma _{i}\tilde{W}_{\Delta
t}^{i}-\frac{1}{2}\alpha _{i}\sigma _{i}^{2}\Delta t). 
\]%
Conditional on the realizations of $X_{1-\Delta t}$ and $M_{1-\Delta
t}^{1/\alpha },$ $\bar{X}$ and $\bar{Y}$ have respective underlying
conditional variances of $\sigma _{0}^{2}\Delta t$ and $\sigma
_{i}^{2}\Delta t,$ as $\sigma _{i}^{M}/\alpha _{i}=\sigma _{i}.$ So the
corresponding regression coefficient $\kappa $ for $\bar{X}$ on $\bar{Y}$ is%
\[
\frac{\alpha ^{2}/(\alpha _{i}\sigma _{i})^{2}}{\alpha ^{2}/(\alpha
_{i}\sigma _{i})^{2}+1/\sigma _{0}^{2}}=\frac{1/(\sigma _{i})^{2}}{1/(\sigma
_{i})^{2}+1/\sigma _{0}^{2}}=\frac{p_{i}}{p_{i}+p_{0}}=\kappa _{1}^{i}, 
\]%
having cancelled $\Delta t>0$ from numerator and denominator; this remains
constant as $\Delta t$ varies, and so also in the passage as $\Delta
t\rightarrow 0.$ Also%
\[
1/\bar{p}=\frac{\Delta t}{1/\sigma _{i}^{2}+1/\sigma _{0}^{2}}, 
\]%
so $\bar{K}_{\alpha }=\bar{K}_{\alpha }(\Delta t)\rightarrow 1$ as $\Delta
t\rightarrow 0.$ Finally, since $Y_{t}^{i}=Z_{t}^{i}M_{t}^{i}=f^{i}X_{t}^{%
\alpha _{i}}M_{t}^{i},$%
\[
\bar{T}=(Y_{1}^{i}/f^{i})^{1/\alpha }=X_{1}M_{1}^{1/\alpha }=\bar{X}\bar{Y}, 
\]%
and so, conditioning on $Y_{1}^{i}=y,$ and setting $k=f^{1-\kappa },$ 
\[
\mathbb{E}[Z_{1}^{i}|\bar{T}]=\mathbb{E}[f^{i}X_{1}^{\alpha }|\bar{T}]=f^{i}(%
\bar{T})^{\alpha \kappa }=f^{i}((Y_{1}^{i}/f^{i})^{1/\alpha })^{\alpha
\kappa }=f^{i}(y/f^{i})^{\kappa }=ky^{\kappa }.\qquad \square 
\]

\bigskip

Below the deterministic functions $\tilde{\beta}^{i}$ factor into $\tilde{%
\beta}_{\text{indiv}}^{i},\tilde{\beta}_{\text{agg}}$ to anticipate $m$-fold
versions which `separate' individual and aggregate effects of agents.

\bigskip

\noindent \textbf{Lemma 2 (Time-}$t$ \textbf{conditional law of the
valuation of }$Z_{1}$\textbf{, given observation }$Y_{t}$ -- \textbf{for }$%
m=1$\textbf{). }\textit{Conditional on }$Y_{t}^{i}=y,$\textit{\ the time-}$t$
\textit{distribution of the time-}$1$ \textit{valuation }$\mathbb{E}%
[Z_{1}|Y_{1},\mathcal{G}_{1}]$\textit{\ is that of}%
\[
k\tilde{\beta}^{i}y^{\kappa }\hat{Z}_{t}:=k\tilde{\beta}_{\text{indiv}}^{i}%
\tilde{\beta}_{\text{agg}}y^{\kappa }\hat{Z}_{t}, 
\]%
\textit{with }$k=k_{1}^{i}$ \textit{as in Lemma 1, and:}\newline
\noindent (i)%
\begin{eqnarray*}
\kappa &=&\kappa _{1}^{i},\text{ }\tilde{\beta}_{\text{indiv}}^{i}:=(\mu
_{t}^{0}(\alpha _{i})\mu _{t}^{i})^{\kappa },\text{ }\tilde{\beta}_{\text{agg%
}}:=\mu (\kappa ,\alpha _{i}^{2}\tilde{\sigma}_{0i}^{2})\text{,} \\
\mu _{t}^{i} &:&=\mu (\alpha _{i},\alpha _{i}^{2}\tilde{\sigma}_{i}^{2})%
\text{ and }\mu _{t}^{0}(\alpha _{i})=\mu (\alpha _{i},\tilde{\sigma}%
_{0}^{2});
\end{eqnarray*}%
\noindent (ii)\textit{\ }$\hat{Z}_{t}$\textit{\ log-normal, its underlying
mean-zero normal of variance} $\hat{\sigma}_{t}^{2}=\kappa \alpha _{i}^{2}%
\tilde{\sigma}_{0i}^{2}$\textit{.}\newline
\textit{In particular, this time-}$t$ \textit{distribution} \textit{has mean
given by}%
\[
\mathbb{E}[Z_{1}|Y_{t}^{i}=y,\mathcal{G}_{t}]=k\tilde{\beta}_{\text{indiv}%
}^{i}\tilde{\beta}_{\text{agg}}y^{\kappa }. 
\]

\noindent \textbf{Proof. }From (\ref{decomp}) with $M=M^{i}$ and \textit{any}
$\delta >0$ 
\[
M_{t+s}^{\delta }=M_{0}^{\delta }\exp (\delta \sigma _{M}W_{t+s}-\frac{1}{2}%
\delta \sigma _{M}^{2}(t+s))=M_{t}^{\delta }\exp (\delta \sigma _{M}\tilde{W}%
_{s}-\frac{1}{2}\delta \sigma _{M}^{2}s). 
\]%
So, for $s=1-t,$%
\[
M_{1}^{\delta }=\mu _{t}(\delta )M_{t}^{\delta }\exp (\delta \sigma _{M}%
\tilde{W}_{1-t}-\frac{1}{2}\delta ^{2}\sigma _{M}^{2}(1-t)), 
\]%
where the last term has unit-mean and%
\[
\mu _{t}(\delta )=\mu (\delta ,\tilde{\sigma}_{M}^{2})=\mu (\delta ,\alpha
_{i}^{2}\tilde{\sigma}_{i}^{2}). 
\]%
In particular, for $\delta =\kappa =\kappa _{1}^{i}$ (i.e. for $\kappa $ as
in Lemma 1) and $M=M^{i},$%
\[
M_{1}^{\kappa }=\mu _{t}^{i}\cdot M_{t}^{\kappa }\cdot \exp (\kappa \alpha
_{i}\sigma _{i}W_{t}^{i}(1-t)-\frac{1}{2}\kappa ^{2}\alpha _{i}^{2}\sigma
_{i}^{2}(1-t)), 
\]%
where $\mu _{t}^{i}=\mu (\kappa ,\alpha _{i}^{2}\tilde{\sigma}_{i}^{2})=\mu
(\kappa _{1}^{i},\alpha _{i}^{2}\tilde{\sigma}_{i}^{2});$ likewise, for $M=X$
and $\delta =\alpha _{i},$ 
\[
Z_{1}^{i}=f^{i}X_{1}^{\alpha _{i}}=\mu _{t}^{0}(\alpha _{i})\cdot
Z_{t}^{i}\cdot \exp (\alpha _{i}\sigma _{0}\tilde{W}_{1-t}^{0}-\frac{1}{2}%
\alpha _{i}^{2}\sigma _{0}^{2}(1-t)), 
\]%
where $\mu _{t}^{0}(\alpha _{i})=\mu (\alpha _{i},\tilde{\sigma}_{0}^{2}).$
Combining, as $Y_{t}^{i}=Z_{t}^{i}M_{t}^{i},$ for any $\delta >0:$%
\[
(Y_{1}^{i})^{\delta }=(\mu _{t}^{0}(\alpha _{i})\mu _{t}^{i}\cdot
Z_{t}^{i}M_{t}^{i})^{\delta }\cdot \exp (\alpha _{i}\delta \sigma _{0}\tilde{%
W}_{1-t}^{0}-\delta \alpha _{i}^{2}\sigma _{0}^{2}(1-t)/2)\cdot \exp (\delta
\alpha _{i}\sigma _{i}\tilde{W}_{1-t}^{i}-\delta \alpha _{i}^{2}\sigma
_{i}^{2}(1-t)/2). 
\]%
But%
\[
\delta \lbrack \alpha _{i}\sigma _{0}\tilde{W}_{1-t}^{0}+\alpha _{i}\sigma
_{i}\tilde{W}_{1-t}^{i}]=\delta \alpha _{i}[\sigma _{0}W_{t}^{0}(1-t)+\sigma
_{i}\tilde{W}_{1-t}^{i}] 
\]%
has variance $\delta ^{2}\alpha _{i}^{2}\tilde{\sigma}_{0i}^{2},$ where $%
\tilde{\sigma}_{0i}^{2}=[\sigma _{0}^{2}+\sigma _{i}^{2}](1-t).$ So taking%
\[
\hat{Z}_{t}(\delta ):=\exp \left( \delta \alpha _{i}[\sigma _{0}\tilde{W}%
_{1-t}^{0}+\sigma _{i}\tilde{W}_{1-t}^{i}]-\frac{1}{2}\delta ^{2}\alpha
_{i}^{2}\tilde{\sigma}_{0i}^{2}\right) , 
\]%
which has unit-mean and variance $\delta ^{2}\alpha _{i}^{2}\tilde{\sigma}%
_{0i}^{2}$, gives%
\begin{equation}
(Y_{1}^{i})^{\delta }=(\mu _{t}^{0}\mu _{t}^{i}\cdot Y_{t}^{i})^{\delta }\mu
(\delta ,\alpha _{i}^{2}\tilde{\sigma}_{0i}^{2})\hat{Z}_{t}(\delta ).
\label{beta-power}
\end{equation}%
In particular, condition on $Y_{t}^{i}=y,$ take $\delta =\kappa $ above, and
set%
\[
\hat{Z}_{t}:=\hat{Z}_{t}(\kappa ); 
\]%
then, by (\ref{beta-power}) and with $k$ as in Lemma 1 above, the
conditional time-$t$ distribution of $\mathbb{E}[Z_{1}|Y_{1},\mathcal{G}%
_{1}] $ is that of the variable $Z_{t}^{\text{est}}$ given by%
\[
Z_{t}^{\text{est}}:=k(Y_{1}^{i})^{\kappa }=k(\mu _{t}^{0}\mu _{t}^{i}\cdot
y)^{\kappa }\mu (\kappa ,\alpha _{i}^{2}\tilde{\sigma}_{0i}^{2})\hat{Z}_{t}.%
\text{ }\square 
\]

\noindent \textbf{Proof of Theorem 1.} Below we suppress reference to the
unique agent $i.$ We condition on the event $\theta _{-}^{\text{pub}}(t)\leq
t<s<\theta _{+}^{\text{pub}}(t),$ i.e. that there has been no subsequent
disclosure in $(t,s].$ Denote by $\gamma _{t}$ the c\`{a}dl\`{a}g censor
assumed in Theorem 1. For $t\leq u\leq s,$ let $ND_{u}(\gamma )$ denote the
event that, at time $u,$ either $\Delta N(u)=0$ or the agent observes $Y_{s}$
to be below $\gamma ,$ and let $\tilde{\gamma}_{u}$ be the time-$u$
evaluation of the random variable $\mathbb{E}[Z_{1}|Y_{1},\mathcal{G}_{1}]$;
then%
\[
\tilde{\gamma}_{u}=\mathbb{E}[Z_{1}|ND_{u}(\gamma _{u}),\mathcal{G}%
_{t}^{+}]. 
\]%
As in \S 2.3, by the Indifference Principle (cf. \cite[\S 11 (Appendix 8)]%
{GieOa}), the unique cutoff value $\gamma _{u}$ for $Y_{u}$ and the time-$u$
evaluation $\tilde{\gamma}_{u}$ of $\mathbb{E}[Z_{1}|Y_{1},\mathcal{G}_{1}]$
are related by 
\[
\tilde{\gamma}_{u}=\mathbb{E}[Z_{1}|ND_{u}(\gamma _{u}),\mathcal{G}_{t}^{+}]=%
\mathbb{E}[Z_{1}|Y_{u}=\gamma _{u},\mathcal{G}_{t}^{+}]=k\tilde{\beta}%
_{u}\gamma _{u}^{\kappa }, 
\]%
with $\tilde{\beta}_{u}=\tilde{\beta}_{u}^{i}$ denoting $\tilde{\beta}^{i}$
evaluated at $u,$ as in Lemma 2 above, so that 
\[
\tilde{\gamma}_{u}=k\tilde{\beta}_{u}\gamma _{u}^{\kappa },\text{ or }\log 
\tilde{\gamma}_{u}=\kappa \log \gamma _{u}+\log k\tilde{\beta}_{u}. 
\]%
For $u>t,$ put $\hat{\gamma}_{u}:=\tilde{\gamma}_{u}/\tilde{\gamma}_{t}$ (so
that $\hat{\gamma}_{t}=1$), thus rescaling $\mathbb{E[E}[Z_{1}|Y_{1},%
\mathcal{G}_{1}]|\mathcal{G}_{t}^{+}],$ the time-$t$ valuation of $\mathbb{E}%
[Z_{1}|Y_{1},\mathcal{G}_{1}],$ to unity; we now work as though $\tilde{%
\gamma}_{t}=1.$ Let the corresponding time-$u$ valuation be the random
variable $\hat{Z}_{u}$ of Lemma 2. The underlying zero-mean normal random
variable of the Lemma has variance $\hat{\sigma}_{u}^{2}:=\kappa ^{2}\tilde{%
\sigma}_{0i}^{2}(u)$. In the notation of Lemma 2, with $q_{ts}:=(s-t)\lambda
_{t}$ the formula of Prop. 1 gives 
\begin{eqnarray*}
(1-q_{ts})(1-\hat{\gamma}_{s})+o(s-t) &=&q_{ts}\int_{z_{s}\leq \hat{\gamma}%
_{s}}(z_{s}-\hat{\gamma}_{s})\mathrm{d}\mathbb{Q}(\hat{Z}_{s}\leq z_{s}|%
\mathcal{G}_{t},\tilde{\gamma}_{t}=1) \\
&=&-q_{ts}\int_{z_{s}\leq \hat{\gamma}_{s}}\mathbb{Q}(\hat{Z}_{s}\leq z_{s}|%
\mathcal{G}_{t},\tilde{\gamma}_{t}=1)\mathrm{d}z_{s}.
\end{eqnarray*}%
Dividing by $-q_{ts}$ and rearranging the differential term, by the
Black-Scholes formula 
\begin{eqnarray*}
&&-(1-q_{ts})\frac{1}{\lambda _{t}}\frac{\hat{\gamma}_{s}-1}{(s-t)}%
=\int_{z_{s}\leq \tilde{\gamma}_{s}}\mathbb{Q}(\hat{Z}_{s}\leq z_{s}|%
\mathcal{G}_{t},\tilde{\gamma}_{t}=1)\mathrm{d}z_{s} \\
&=&\hat{\gamma}_{s}\Phi \left( \frac{\log (\hat{\gamma}_{s})+\frac{1}{2}\hat{%
\sigma}^{2}(1-t)}{\hat{\sigma}\sqrt{1-t}}\right) -\Phi \left( \frac{\log (%
\hat{\gamma}_{s})-\frac{1}{2}\hat{\sigma}^{2}(1-t)}{\hat{\sigma}\sqrt{1-t}}%
\right) ,
\end{eqnarray*}%
up to $o(s-t)/[(s-t)\lambda _{t}].$ Rearranging once more and, using the
abbreviating notation $\hat{\sigma}_{t}^{2}$ for the variance,%
\[
(1-q_{ts})\frac{1}{\lambda _{t}}\frac{\hat{\gamma}_{s}-1}{(s-t)}=-\hat{\gamma%
}_{s}\Phi \left( \frac{\log (\hat{\gamma}_{s})+\frac{1}{2}\hat{\sigma}%
_{t}^{2}}{\hat{\sigma}_{t}}\right) +\Phi \left( \frac{\log (\hat{\gamma}%
_{s})-\frac{1}{2}\hat{\sigma}_{t}^{2}}{\hat{\sigma}_{t}}\right) +\frac{o(s-t)%
}{(s-t)\lambda _{t}}. 
\]%
Because $y_{u}$ is c\`{a}dl\`{a}g, so also are $\tilde{\gamma}_{u}$ and $%
\hat{\gamma}_{u};$ so the terms on the right have a limit as $s\downarrow t.$
As $q_{ts}\rightarrow 0$, the function $\hat{\gamma}(u)$ is seen to be
right-differentiable at $t,$ and, since $\Phi (-u)=1-\Phi (u),$%
\begin{equation}
\frac{1}{\lambda _{t}}\hat{\gamma}^{\prime }(t)=-[2\Phi (\hat{\sigma}%
_{t}/2)-1],  \label{gamma-hat}
\end{equation}%
or equivalently, recalling that $\hat{\sigma}_{t}^{2}=\alpha _{i}^{2}(\sigma
_{0}^{2}+\sigma _{i}^{2})(1-t),$%
\begin{equation}
\frac{\tilde{\gamma}^{\prime }(t)}{\tilde{\gamma}(t)}=-\lambda _{t}[2\Phi (%
\hat{\sigma}_{t}/2)-1].  \label{gamma-tilde}
\end{equation}

Now unfixing $t,$ we permit $t$ to vary over an interval during which there
is no disclosure. Solving the differential equation, by integrating from the
last disclosure date $\theta _{-}(t)\geq 0$ to the date $t,$ and
conditioning on $t$ being prior to the next disclosure $\theta _{+}(t),$
gives the following:%
\begin{eqnarray*}
\log \left( \tilde{\gamma}(t)/\tilde{\gamma}(\theta _{-}(t))\right)
&=&-\int_{\theta _{-}}^{t}\lambda _{u}[2\Phi (\hat{\sigma}_{u}/2)-1]du, \\
\tilde{\gamma}(t) &=&\tilde{\gamma}(\theta _{-}(t))\exp \left( -\int_{\theta
_{-}}^{t}\lambda _{u}[2\Phi (\hat{\sigma}_{u}/2)-1]du\right) .
\end{eqnarray*}%
Note that as $\hat{\sigma}_{u}\geq 0,$ the factor $[2\Phi (\hat{\sigma}%
_{s}/2)-1]$ is non-negative.

So the conditional time-$t$ evaluation of $\mathbb{E}_{1}[Z_{1}|Y_{1}]:=%
\mathbb{E}[Z_{1}|Y_{1},\mathcal{G}_{1}]$ is given explicitly by%
\[
\tilde{\gamma}_{t}=\mathbb{E}[\mathbb{E}_{1}[Z_{1}|Y_{1}]|ND_{t}(\gamma
_{t}),\mathcal{G}_{t}]=\mathbb{E}[Z_{1}|ND_{t}(\gamma _{t}),\mathcal{G}_{t}]=%
\hat{\gamma}(t)\mathbb{E}[Z_{1}|Y_{\theta },\mathcal{G}_{\theta }], 
\]%
for $\theta =\theta _{-}(t)$ (and with $t$ unfixed), where now%
\[
\hat{\gamma}(t):=\exp \left( -\int_{\theta _{-}}^{t}\lambda _{u}[2\Phi (\hat{%
\sigma}_{u}/2)-1]du\right) , 
\]%
by abuse of notation, as this function satisfies (\ref{gamma-hat}), but with 
$\hat{\gamma}(\theta _{-}(t))=1$.

To obtain the explicit form, apply Lemma 2 with $\theta =\theta _{-}$ to give%
\[
\mathbb{E}[Z_{1}|Y_{\theta },\mathcal{G}_{\theta }]=k\beta _{\theta
}Y_{\theta }^{\kappa }, 
\]%
where, re-instating the index $i,$%
\[
\kappa =\kappa _{1}^{i}=p_{i}/(p_{0}+p_{i}),\text{\qquad\ }%
k=k_{1}^{i}=f_{i}^{1-\kappa _{1}^{i}},\text{\qquad }\beta _{t}=(\mu
_{t}^{0}\mu _{t}^{i})^{\kappa _{1}^{i}}\mu (\kappa _{1}^{i},\tilde{\sigma}%
_{0i}^{2}), 
\]%
with $\mu _{t}^{i}:=\mu (\alpha _{i},\alpha _{i}^{2}\tilde{\sigma}_{i}^{2})$
and $\mu _{t}^{0}(\alpha _{i})=\mu (\alpha _{i},\tilde{\sigma}_{0}^{2}).$
Finally, the cutoff for $Y_{u}^{i}$ is given explicitly by Lemma 2 as 
\[
\gamma _{u}^{i}=(\tilde{\gamma}_{u}^{i}/(k_{1}^{i}\beta _{u}^{i}))^{1/\kappa
_{1}^{i}}. 
\]

\subsection{Proof of Theorem 1$_{m}$}

This section\textbf{\ }is devoted to a proof of Theorem 1$_{m},$ which is
like Theorem 1, but more intricate in its details on account of an
application of a result from \cite[Th. 14.2 (Appendix 7)]{GieOa}, Theorem M
below. Appropriate substitutions (of the parameter values used here for the
parameter values used there) are needed; justification of these is routine,
but cumbersome, so shown as a tabulation in\textbf{\ this the proposed} the
arXiv version only; their basis comes from some calculations deferred to 
\textbf{\S }5.4. In view of similarities, as well as differences of context,
between the present and the source paper, we follow the convention of 
\textbf{\S }5.2 of \textit{overbarring} variables cited from \cite[Th. 14.2
(Appendix 7)]{GieOa}. This allows Theorem M to be read as applicable in
either of the two contexts according as parameters are over-barred, or not.
We note that $\bar{\lambda}=1/\lambda $ -- the $\lambda $ variables of the
two papers are reciprocals, by Prop. 1.

\bigskip

\textbf{Theorem M (Multi-firm Cutoff Equations}, \cite[Th. 14.2 (Appendix 7)]%
{GieOa}\textbf{).} \textit{In the setting of this section, after rescaling
so that }$Y_{t}^{i}=1$\textit{, with observations }$Y_{s}^{i}$\textit{\ of
time }$s>t$ \textit{replaced by their re-scaled versions }$\tilde{Y}_{s}^{i}$%
\textit{, the simultaneous conditional Bayes equations determining the
cutoffs for }$\tilde{Y}_{s}^{i}$ \textit{may be reduced to a non-singular
system of linear equations relating the log-cutoffs to the hypothetical
cutoffs }$g_{i}$\textit{\ defined below. Furthermore, the unique solution
for the disclosure cutoff }$\tilde{y}^{i}$ \textit{for the observation of }$%
\tilde{Y}_{s}^{i}$\textit{\ is given by }%
\begin{equation}
\log \tilde{y}^{i}=\frac{\log g_{i}}{\alpha _{i}\kappa _{-i}}+\frac{1}{%
\kappa _{0}}\left( \frac{\kappa _{1}}{\alpha _{1}\kappa _{-1}}\log g_{1}+%
\frac{\kappa _{2}}{\alpha _{2}\kappa _{-2}}\log g_{2}+...+\frac{\kappa _{m}}{%
\alpha _{m}\kappa _{-m}}\log g_{m}\right) ,  \label{multi-Dye-Fmla}
\end{equation}%
\textit{where}%
\begin{eqnarray*}
g_{i} &=&g_{i}(s)=\hat{\gamma}_{\text{L N}}(\tilde{\lambda}_{i},\alpha
_{i}\kappa _{i}\tilde{\sigma}_{0i}\sqrt{1-\tilde{\rho}_{i}^{2}})L_{-i}\text{
with }\tilde{\lambda}_{i}=\lambda _{i}(s),\text{ the }N_{s}^{i}\text{
intensity,} \\
L_{-i} &=&L_{-i}(s)=\exp \left( \frac{\alpha _{i}(m-1)+\alpha _{i}(\alpha
_{i}-1)}{2(\tilde{p}-\tilde{p}_{i})}\right) \exp \left( -\frac{m\alpha
_{i}+\alpha _{i}(\alpha _{i}-1)}{2\tilde{p}}\right) , \\
&&(\text{the `amended mean' -- an adjustment coefficient for the cutoff),}
\end{eqnarray*}%
\textit{and where:}\newline
\noindent $\hat{\gamma}_{\text{LN}}(\lambda ,\sigma )$\textit{\ denotes the
unique solution of the following equation in }$y:$\textit{\ }%
\begin{equation}
(1-y)=\lambda H_{\text{LN}}(y,\sigma ),  \label{implicit-Dye}
\end{equation}

\noindent $\kappa _{i}=\bar{p}_{i}/\bar{p}$\textit{\ (the standard
regression coefficient),}\newline
\noindent $\kappa _{-i}=\bar{p}_{i}/(\bar{p}-\bar{p}_{i})$ \textit{(removing
agent-}$i$\textit{'s contribution from the aggregate precision),}\newline
\noindent $1-\tilde{\rho}_{i}^{2}$ \textit{is the partial covariance of }$%
\tilde{w}_{i}$ \textit{on the remaining variates }$\tilde{w}_{j}$\textit{,
with }$\tilde{w}_{j}(t):=\sigma _{0}\tilde{W}_{1-t}^{0}+\sigma _{i}\tilde{W}%
_{1-t}^{j}.$

\bigskip

\noindent \textbf{Proof strategy. }The results of \cite{GieOa} concern a
two-period model in which there is an initial time (taken here to be a fixed
time $t<1$), a second `interim date' (taken here to be a time $s$ with $%
t<s<1 $), and a terminal date of $1$, as here. Furthermore, that model
refers to random variables, whose realizations describe values which are to
be disclosed at the (mandatory) disclosure terminal date. Agent $i$ in \cite%
{GieOa} has at the interim date (time $s,$ here) a positive probability of
receiving a private, noisy observation $\bar{T}_{i}=\bar{X}\bar{Y}_{i}$ with 
$\bar{X}$ the state random variable and $\bar{Y}_{i}$ the noise$;$ the
random variables are independent, log-normally distributed, with underlying
normals having zero-mean and precisions $\bar{p}_{0}$ and $\bar{p}_{i},$
respectively. The probability of observation is described by the odds $\bar{%
\lambda}_{i}$. The agent seeks to maximize the interim-date expected value
of $Z^{i}:=f_{i}\bar{X}^{\alpha _{i}},$ by selecting a unique cutoff $\bar{%
\gamma}^{i}$ for $\bar{T}_{i}.$ The model may be notationally summarized as $%
\mathfrak{M}^{\bar{n}}(Z|\bar{T}_{1},...,\bar{T}_{\bar{n}};\bar{\lambda}%
_{1},...,\bar{\lambda}_{\bar{n}}),$ with $\bar{n}$ the number of agents. The
value of $\bar{\gamma}^{i}$ is unique and is described in Theorem M by
suitable aggregation of cutoffs $\hat{\gamma}_{i}$ obtained by reference to $%
\bar{n}$ independent simpler single-agent models $\mathfrak{M}^{1}(Z_{1}^{%
\text{hyp-}i}|\bar{T}_{i}^{\text{hyp}};\bar{\lambda}_{i}),$ the solitary
agents $i$ being the \textit{hypothetical} agents (corresponding to the
original agents $i$); the original correlations between agents are removed,
by way of the Schur complement -- specifically via a partial covariance $%
\bar{\rho}_{i}$ (measuring covariance of the $i$-th agent on the others, as
described in the theorem; see \cite[Note 4.27, p.120]{BinF}, cf. \cite[Ch. 27%
]{KenS-2}, \cite[\S \S 46.26-28]{KenS-3}). Theorem M\ prescribes $\hat{\gamma%
}_{i}$ for the common-sized (cf. \S 2.1.2) hypothetical observation variate:%
\[
\bar{T}_{i}^{\text{hyp}}:=(\bar{T}_{i})^{\alpha _{i}\bar{\kappa}_{i}\sqrt{1-%
\bar{\rho}_{i}^{2}}}=(\bar{X}\bar{Y}_{i})^{\alpha _{i}\bar{\kappa}_{i}\sqrt{%
1-\bar{\rho}_{i}^{2}}}, 
\]%
i.e. with precision parameter altered by the factor $\alpha _{i}\bar{\kappa}%
_{i}\sqrt{1-\bar{\rho}_{i}^{2}}$.

This result comes from employing, in the spirit of \textbf{\S }5.2, a
monotone single-variable conditional regression function, e.g. $\tilde{m}%
(t_{1}|\bar{\gamma}^{2},\bar{\gamma}^{3},...):=$ $\mathbb{E}[Z_{1}^{i}|\bar{T%
}_{1}=t_{1}\&(\forall j>1)(\bar{T}_{j}=$ $\bar{\gamma}^{j})],$ to find the
cutoff for the corresponding conditional valuation 
\[
\tilde{Z}_{1}:=\mathbb{E}[Z_{1}^{i}|\bar{T}_{1}=t_{1}\&(\forall j>1)(\bar{T}%
_{j}=\bar{\gamma}^{j})]. 
\]%
The characterizing equation reduces to an equivalent unconditional one,
where all but one of the noisy observations is absent.

The proof in outline runs like this.

(i) Apply Theorem M. To achieve the format $\bar{T}_{i}=\bar{X}\bar{Y}_{i}$
replace $Y_{t}^{i}$ by 
\[
\tilde{Y}_{t}^{i}=(Y^{i})^{1/\alpha _{i}}. 
\]%
Fix times $t,s$ with $t<s;$ assume non-disclosure throughout $(t,s];$
construct random variables, which at time $s$ describe outcomes of time $1;$
for these compute appropriate statistics; pass to the hypothetical
processes. At this point the latter are defined \textit{implicitly} only.

Much of this part of the proof is deferred to \textbf{\S }5.4 Lemmas 1$_{m}$
and 2$_{m}$ (versions of Lemmas 1 and 2 for general $m$), where we compute
the constants $k_{m}^{i}$ and deterministic functions $\beta _{t}^{i}$ such
that for $\kappa _{i}=\kappa _{m}^{i}$%
\[
\mathbb{E}[Z_{1}^{i}|Y_{t}=y_{t},\mathcal{G}_{t}]=k_{m}^{i}\beta
_{t}^{i}y_{1t}^{\kappa _{1}}...y_{mt}^{\kappa _{i}}. 
\]

(ii) Obtain \textit{explicit} formulas for $y_{s}^{\text{hyp}}$ by
variational analysis: compute the dynamics of the valuation cutoffs $\gamma
_{t}^{i}$ of the `estimators' $\hat{Z}_{t}^{\text{hyp }i}=\mathbb{E}[Z_{1}^{%
\text{hyp-}i}|\mathcal{G}_{t}]$ from the dynamics of the hypothetical
processes. Being hypothetical models with state and observation processes
identical, these observation and valuation cutoffs are the same as $y_{s}^{%
\text{hyp}}$.

(iii) First Inversion: substitute $y_{s}^{\text{hyp}}$ into Theorem M to
obtain cutoffs $\bar{\gamma}_{s}^{i}=\tilde{y}_{s}^{i}$ for $\tilde{Y}%
_{s}^{i}$ from the vector of observation cutoffs $y_{s}^{\text{hyp}}$ of the
hypothetical agents $j$.

(iii) Second Inversion: Reverse-engineer from the $\tilde{y}_{s}^{i}$ cutoff
dynamics the dynamics for $Y_{t}^{i}=(\tilde{Y}_{t}^{i})^{\alpha _{i}},$
using $\log y_{s}^{i}=\alpha _{i}\log \tilde{y}_{s}^{i}.$

\bigskip

\noindent \textbf{Substitutions justified. }Taking $\hat{f}%
^{i}=(f^{i})^{1/\alpha _{i}}$ and $\tilde{M}_{s}^{i}:=\left(
M_{s}^{i}\right) ^{1/\alpha _{i}},$ 
\[
\tilde{Y}_{t}^{i}:=(Y_{t}^{i})^{1/\alpha
_{i}}=(Z_{t}^{i}M_{t}^{i})^{1/\alpha _{i}}=\hat{f}^{i}\left( X_{t}\right) 
\tilde{M}_{t}^{i}, 
\]%
(cf. Lemmas 1 and 2 of \textbf{\S }5.2), so $\tilde{Y}_{1}^{i}=\hat{f}%
^{i}X_{1}\tilde{M}_{1}^{i}.$

Since $\tilde{M}_{t}^{i}$ has underlying Wiener volatility $\sigma
_{i}^{M}/\alpha _{i}=\sigma _{i},$ we compute parameter values as tabulated
below.\renewcommand{\arraystretch}{1.25}%
\[
\begin{tabular}{|l|l|}
\hline
$\overline{\text{There}}\text{ (barred)}$ & $\text{Here}$ \\ \hline
$\bar{\sigma}_{0}$ & $\tilde{\sigma}_{0}:=\tilde{\sigma}_{0}(t)=\sigma
_{0}(1-t)^{1/2}$ \\ \hline
$\bar{\sigma}_{i}$ & $\tilde{\sigma}_{i}\text{ }i=0,1,...,m,$ \\ \hline
$\bar{\sigma}_{0i}^{2}:=\bar{\sigma}_{0}^{2}+\bar{\sigma}_{i}^{2}$ & $\tilde{%
\sigma}_{0i}^{2}=\tilde{\sigma}_{0}^{2}+(\tilde{\sigma}_{i})^{2}$ \\ \hline
$\bar{p}_{i},\bar{p}$ & $%
\begin{tabular}{|l|}
\hline
$p_{i}=1/\sigma _{i}^{2},\text{ }p=\sum_{i=0}^{m}1/\sigma _{i}^{2}$ \\ \hline
$\tilde{p}_{i}=p_{i}(1-t),\text{ }\tilde{p}=p(1-t)$ \\ \hline
\end{tabular}%
$ \\ \hline
$\bar{\kappa}_{i}=\bar{p}_{i}/\bar{p}$ & $\tilde{\kappa}_{i}=\tilde{p}_{i}/%
\tilde{p}=p_{i}/p=\kappa _{i}\text{ (constants)}$ \\ \hline
$T_{i}=e^{\bar{\sigma}_{0i}\bar{w}_{i}-\frac{1}{2}\bar{\sigma}_{0i}^{2}}$ & $%
\tilde{Y}_{1}^{i}=\tilde{Y}_{t}^{i}\left( e^{\sigma _{0}\tilde{W}%
_{t}^{0}(1-t)-\frac{1}{2}\tilde{\sigma}_{0}^{2}}\right) \left( e^{\sigma _{i}%
\tilde{W}_{t}^{i}(1-t)-\frac{1}{2}\tilde{\sigma}_{i}{}^{2}}\right) $ \\ 
\hline
$\bar{w}_{i}$ & $\tilde{w}_{i}:=\sigma _{0}\tilde{W}_{t}^{0}(1-t)+\sigma _{i}%
\tilde{W}_{t}^{i}(1-t)$ \\ \hline
$\bar{\rho}_{i}$ & $\tilde{\rho}_{i}$ \\ \hline
$X^{\alpha _{i}}$ & $Z_{1}^{i}:=f^{i}\cdot X_{1}^{\alpha _{i}}$ \\ \hline
$\bar{\sigma}_{\text{hyp,}i}$ & $\sigma _{\text{hyp,}i}:=\alpha _{i}\kappa
_{i}\tilde{\sigma}_{0i}\sqrt{1-\tilde{\rho}_{i}^{2}}$ \\ \hline
\end{tabular}%
\]%
\renewcommand{\arraystretch}{1}\newline

\noindent \textbf{Proof of Theorem 1}$_{m}.$We proceed stepwise.

\noindent \textbf{1. (Hypothetical cutoff dynamics). }We consider $s,t$ with 
$\theta =\theta _{-}^{\text{pub}}\leq t<s<\theta _{+}^{\text{pub}}\leq 1.$
We apply Theorem M above (with $t$ as the ex-ante date and $s$ the interim
date) to an agent $i.$ Theorem M sets the observation cutoffs for $\tilde{Y}%
_{u}^{i}$ in terms of the cutoffs $y_{i}^{\text{hyp}}(u)$ of a
correspondingly defined `hypothetical' observer, the latter being defined
implicitly via equation (\ref{implicit-Dye}). As in \textbf{\S }5.2, we
perform a variational analysis to derive explicitly the cutoffs $\hat{z}%
_{i}(u)$ of the corresponding hypothetical observation.

Since Theorem M applies to variables that have been common-sized to unity at
the ex-ante date $t$, put $G_{i}(u)=\hat{z}_{i}(u)/\hat{z}_{i}(t),$ so that $%
G_{i}(t)=1$ for each $i.$ Then, corresponding to a common-sized process,
there is a hypothetical process ($g$-process) with adjusted value $%
G_{i}(u)L_{-i}(u)$ as at time $u,$ and with a hypothetical volatility per
unit time of 
\[
\tilde{\sigma}_{\text{hyp,}i}=\sigma _{\text{hyp,}i}(u):=\alpha _{i}\kappa
_{i}\tilde{\sigma}_{0i}\sqrt{1-\tilde{\rho}_{i}^{2}}. 
\]%
The solitary hypothetical process, observed intermittently by agent $i,$ is
now subjected to the variational analysis of \textbf{\S }5.2, as follows.

As in \S 3, by the Indifference Principle of \cite[\S 15 (Appendix 8)]{GieOa}
applied at time $t,$ the unique cutoff value $\tilde{y}_{is}$ for $\tilde{Y}%
_{s}^{i}$ and the time-$s$ evaluation of $\mathbb{E}%
[Z_{1}^{i}|Y_{1}^{1},...,Y_{1}^{m}]$ are related by%
\begin{eqnarray*}
\mathbb{E}[Z_{1}^{i}|(\forall j)ND_{j}(\tilde{y}_{jt}),\mathcal{G}_{t}] &=&%
\mathbb{E}[Z_{1}^{i}|(\forall j\neq i)ND_{j}(\tilde{y}_{jt}),\bar{T}_{i}=%
\tilde{y}_{it},\mathcal{G}_{t}]=\mathbb{E}[Z_{1}^{i}|(\forall j)[\bar{T}_{j}=%
\tilde{y}_{j}],\mathcal{G}_{t}] \\
&=&k_{m}^{i}\beta _{t}^{i}\tilde{y}_{1t}^{\alpha \kappa _{1}}...\tilde{y}%
_{mt}^{\alpha \kappa _{m}},
\end{eqnarray*}%
for $\alpha =\alpha _{i},$ and $\kappa _{i}=p_{i}/p$ as above; here the last
regression formula is quoted from Lemma 2$_{m}$ of \S 5.4 below (cf. \cite[%
\S 10.3.3 (Appendix 3)]{GieOa}.)

Dropping subscripts, as in \textbf{\S }5.2, one has by the conditional Bayes
formula (Prop. 1, \textbf{\S }2.3) for the hypothetical output valuation:%
\[
G(t)L(t)-G(s)L(s)=\lambda _{t}(s-t)\int_{z_{1}\leq G(s)L(s)}\mathbb{Q}(%
\mathbb{E}[Z_{1}^{\text{hyp}}|Y_{s}^{\text{hyp}},\mathcal{G}_{t}^{+}]\leq
z_{1}|\mathcal{G}_{t}^{+})dz_{1}+o(s-t), 
\]%
where $Z_{1}^{\text{hyp}}=Y_{1}^{\text{hyp}},$ by definition of the
hypothetical agent. Now argue as in \textbf{\S }5.2 with $G(t)L(t)$ in place
of $\tilde{\gamma}(t)$ to deduce the analogue of (\ref{gamma-tilde}).
Rearranging, and using the Black-Scholes put formula, just as in \S 5.2, we
obtain to within $o(s-t)/[(s-t)\lambda _{t}]$%
\begin{eqnarray*}
&&-\frac{1}{\lambda _{t}}\cdot \frac{G(s)L(s)-G(t)L(t)}{(s-t)}=G(s)L(s)\Phi
\left( \frac{\log (G(s)L(s)/G(t)L(t))+\tilde{\sigma}_{\text{hyp}}^{2}/2}{%
\tilde{\sigma}_{\text{hyp}}}\right) \\
&&-G(t)L(t)\Phi \left( \frac{\log (G(s)L(s)/G(t)L(t)))-\tilde{\sigma}_{\text{%
hyp}}^{2}/2)}{\tilde{\sigma}_{\text{hyp}}}\right) .
\end{eqnarray*}%
Now $G(u)/G(t)=\hat{z}_{u}/\hat{z}_{t}$ is a c\`{a}dl\`{a}g process (as a
function of $u$, because each $\tilde{y}_{ju}$ is, and the equations
connecting the log-cutoffs and the hypothetical log-cutoffs have
non-singular matrix). So the right-hand side has a limiting value as $%
s\downarrow $ $t.$ Hence $G(u)L(u),$ and so $G(u),$ is right-differentiable
at $u=$ $t<1$. Passing to the limit (as in \textbf{\S }5.2)

\[
-\frac{1}{\lambda _{t}}\cdot \left. \frac{d}{du}\left[ G(u)L(u)\right]
\right\vert _{u=t}=G(t)L(t)[2\Phi (\tilde{\sigma}_{\text{hyp}}/2)-1]. 
\]%
Performing the differentiation, we obtain the following%
\begin{eqnarray*}
G^{\prime }(t)L(t)+G(t)L^{\prime }(t) &=&-\lambda _{t}G(t)L(t)[2\Phi (\tilde{%
\sigma}_{\text{hyp}}/2)-1], \\
\frac{G^{\prime }(t)}{G(t)}+\frac{L^{\prime }(t)}{L(t)} &=&-\lambda
_{t}[2\Phi (\tilde{\sigma}_{\text{hyp}}/2)-1].
\end{eqnarray*}%
But $\hat{z}(u)=\hat{z}(t)G(u)$ [and $\hat{z}^{\prime }(u)=\hat{z}%
(t)G^{\prime }(u),$ so $\hat{z}^{\prime }(u)/\hat{z}(u)=G^{\prime }(u)/G(u)$%
], so 
\[
\frac{\hat{z}^{\prime }(t)}{\hat{z}(t)}+\frac{L^{\prime }(t)}{L(t)}=-\nu _{%
\text{hyp}}(t):=-\lambda _{t}[2\Phi (\tilde{\sigma}_{\text{hyp}}/2)-1]. 
\]%
So for the hypothetical agent $i$ we obtain explicitly that 
\[
\log \left( \hat{z}_{i}(t)/\hat{z}_{i}(\theta )\right) +\log \left(
L_{-i}(t)/L_{-i}(\theta )\right) =-\int_{\theta }^{t}\nu _{i\text{hyp}%
}(s)ds, 
\]%
by integrating from the date $\theta =\theta _{-}$ up to any time $t$ prior
to $\theta _{+}$ and re-instating subscripts. So%
\[
\log \left( \hat{z}_{i}(t)L_{-i}(t)\right) =\log \left( \hat{z}_{i}(\theta
)L_{-i}(\theta )\right) -\int_{\theta }^{t}\nu _{i\text{hyp}}(s)ds, 
\]%
using the definition of $\nu _{i\text{hyp}}.$ But $y_{i}^{\text{hyp}}(u)=%
\hat{z}_{i}(u)L_{-i}(u),$ so%
\begin{equation}
\log y_{i}^{\text{hyp}}(u)=\log y_{i}^{\text{hyp}}(\theta )-\int_{\theta
}^{t}\nu _{i\text{hyp}}(s)ds.  \label{GLi}
\end{equation}%
\noindent \textbf{2. (Actual correlated observation dynamics). }We apply the
formula of Theorem M to obtain the cutoffs $\tilde{y}_{t}^{i}$ for $\tilde{Y}%
_{t}^{i}.$ To have common-sizing at time $\theta ,$ set $\hat{\gamma}%
_{i}(t):=\tilde{y}^{i}(u)/\tilde{y}^{i}(\theta )$ for $\theta =\theta
_{-}<t<\theta _{+}.$ Returning to the corresponding actual agent $i,$
substitution from (\ref{GLi}) into (\ref{multi-Dye-Fmla}) gives%
\begin{eqnarray*}
&&\log \hat{\gamma}_{i}(t)=\frac{1}{\alpha _{i}\kappa _{-i}}\left( \log
y_{i}^{\text{hyp}}(\theta )-\int_{\theta }^{t}\nu _{i\text{hyp}}(s)ds\right)
\\
&&+\frac{1}{\kappa _{0}}\left( ...+\frac{\kappa _{j}}{\alpha _{j}\kappa _{-j}%
}\left( \log (y_{j}^{\text{hyp}}(\theta ))-\int_{\theta }^{t}\nu _{j\text{hyp%
}}(s)ds\right) +...\right) .
\end{eqnarray*}%
We rearrange this to separate out two components displayed below. The first,
a continuous \textquotedblleft inter-arrival discounting\textquotedblright\
term, generalizes the single observer case, as a regression-weighted average
over the hypothetical counterparts: 
\begin{eqnarray*}
&&\frac{1}{\alpha _{i}\kappa _{-i}}\int_{\theta }^{t}\nu _{i\text{hyp}}(s)ds+%
\frac{1}{\kappa _{0}}\left( \frac{\kappa _{1}}{\alpha _{1}\kappa _{-1}}%
\int_{\theta }^{t}\nu _{1\text{hyp}}(s)ds+\frac{\kappa _{2}}{\alpha
_{2}\kappa _{-2}}\left( ...\right) +...\right) \\
&=&\frac{1}{\alpha _{i}\kappa _{-i}}\int_{\theta }^{t}\nu _{i\text{hyp}%
}(s)ds+\frac{1}{\kappa _{0}}\left( \frac{\kappa _{1}}{\alpha _{1}\kappa _{-1}%
}\int_{\theta }^{t}\nu _{1\text{hyp}}(s)ds+\frac{\kappa _{2}}{\alpha
_{2}\kappa _{-2}}\int_{\theta }^{t}\nu _{2\text{hyp}}(s)ds+...\right) \\
&=&\int_{\theta }^{t}\nu _{i}(s)ds.
\end{eqnarray*}%
The second, a shift term, arises from adjustment of the initial mean in the
move from actual to hypothetical agent: 
\[
\frac{1}{\alpha _{i}\kappa _{-i}}\log y_{i}^{\text{hyp}}(\theta )+\frac{1}{%
\kappa _{0}}\left( ...+\frac{\kappa _{i}}{\alpha _{i}\kappa _{-i}}\log
y_{i}^{\text{hyp}}(\theta )+...\right) , 
\]%
necessarily equivalent (after factoring through by $\alpha _{i},$ since$%
(\log y^{i}(u))/\alpha _{i}=\log \tilde{y}^{i}(u))$ to $(\log y_{i}(\theta
))/\alpha _{i}$ for $y_{i}(\theta )$ the re-initialization value at time $%
\theta .$ So%
\begin{eqnarray*}
\frac{1}{\alpha _{i}}\log y^{i}(u) &=&\log \tilde{y}^{i}(u)=-\frac{1}{\alpha
_{i}\kappa _{-i}}\int_{\theta }^{t}\nu _{i\text{hyp}}(s)ds \\
&&-\frac{1}{\kappa _{0}}\left( \frac{\kappa _{1}}{\alpha _{1}\kappa _{-1}}%
\int_{\theta }^{t}v_{1\text{hyp}}(s)ds+\frac{\kappa _{2}}{\alpha _{2}\kappa
_{-2}}\int_{\theta }^{t}v_{2\text{hyp}}(s)ds+...\right) .
\end{eqnarray*}%
So%
\begin{eqnarray*}
\log y_{t}^{i} &=&-\frac{1}{\kappa _{-i}}\int_{\theta }^{t}\nu _{i\text{hyp}%
}(s)ds \\
&&-\frac{1}{\kappa _{0}}\left( \kappa _{1}\frac{\alpha _{i}}{\alpha _{1}}%
\frac{1}{\kappa _{-1}}\int_{\theta }^{t}v_{1\text{hyp}}(s)ds+\kappa _{2}%
\frac{\alpha _{i}}{\alpha _{2}}\frac{1}{\kappa _{-2}}\int_{\theta }^{t}v_{2%
\text{hyp}}(s)ds+...\right) .
\end{eqnarray*}%
\noindent \textbf{3. (Actual correlated valuation dynamics). }From Lemma 2$%
_{m}$%
\[
\tilde{\beta}^{i}=\mu (\alpha _{i},\kappa \tilde{\sigma}_{0}^{2})\mu (\kappa
_{1}^{i},\kappa \tilde{\sigma}_{i}^{2}),\text{ and }\tilde{\beta}%
_{m}=\prod\nolimits_{j}\mu (\kappa _{j},\tilde{\sigma}_{0j}^{2}), 
\]%
\[
\mathbb{E}[Z_{1}^{h}|Y_{t}=y_{t},\mathcal{G}_{t}]=k_{m}^{h}\tilde{\beta}%
^{h}\cdot \tilde{\beta}_{m}\cdot y_{1t}^{\kappa _{1}}...y_{mt}^{\kappa
_{m}}, 
\]%
using notation established there. Put%
\[
\tilde{\gamma}_{t}=y_{1t}^{\kappa _{1}}...y_{mt}^{\kappa _{m}}. 
\]%
Take logarithms, substitute for $\kappa _{i}\log y_{it},$ next relabel $j$
for $i$ in the first term and change order of summation in the second, to
obtain%
\begin{eqnarray*}
\log \tilde{\gamma}_{t} &:&=-\sum\nolimits_{i}\frac{\kappa _{i}}{\kappa _{-i}%
}\int_{\theta }^{t}\nu _{i\text{hyp}}(s)ds-\sum\nolimits_{i}\sum\nolimits_{j}%
\frac{\kappa _{i}}{\kappa _{0}}\frac{\alpha _{i}}{\alpha _{j}}\frac{\kappa
_{j}}{\kappa _{-j}}\int_{\theta }^{t}v_{j\text{hyp}}(s)ds \\
&=&-\sum\nolimits_{j}\frac{\kappa _{j}}{\kappa _{-j}}\int_{\theta }^{t}\nu
_{j\text{hyp}}(s)ds-\sum\nolimits_{j}\sum\nolimits_{i}\frac{\kappa _{i}}{%
\kappa _{0}}\frac{\alpha _{i}}{\alpha _{j}}\frac{\kappa _{j}}{\kappa _{-j}}%
\int_{\theta }^{t}v_{j\text{hyp}}(s)ds \\
&=&-\sum\nolimits_{j}\left( 1+\sum\nolimits_{i}\frac{\alpha _{i}}{\alpha _{j}%
}\frac{\kappa _{i}}{\kappa _{0}}\right) \frac{\kappa _{j}}{\kappa _{-j}}%
\int_{\theta }^{t}\nu _{j\text{hyp}}(s)ds.
\end{eqnarray*}%
So%
\[
\mathbb{E}[Z_{1}^{i}|Y_{t}=y_{t},\mathcal{G}_{t}]=k_{m}^{i}\tilde{\beta}%
^{i}\cdot \tilde{\beta}_{m}\cdot \tilde{\gamma}_{t}\cdot g_{\ast
}^{i}[\theta _{-}]. 
\]

\subsection{Extending the two lemmas of \S 5.2}

We extend Lemmas 1 and 2 of \textbf{\S }5.2 to general $m$ to establish:%
\[
\tilde{\gamma}_{t}^{i}=\mathbb{E}[Z_{1}^{i}|Y_{t}=y_{t},\mathcal{G}%
_{t}]=k_{m}^{i}\beta _{t}^{i}y_{1t}^{\kappa _{1}}...y_{mt}^{\kappa _{m}}. 
\]%
We need some additional notation chosen so as to have the general case
appear typographically similar to the $m=1$ case. Treating $m$-vectors $%
y,z\in \mathbb{R}_{+}^{m}$ as functions on $\{1,2,3,...,m\}$, define $y\cdot
z$ (and so $y/z$) and the exponential $y^{z}$ in the pointwise sense; also
write the \textit{product} of the components of $y^{z}$ as 
\[
\langle y^{z}\rangle :=(y_{1}{}^{z_{1}})...(y_{m}{}^{z_{m}}), 
\]%
by analogy with inner products, so that%
\[
\log \langle y^{z}\rangle :=\langle z,\log y\rangle =z_{1}\log
y_{1}+...+z_{m}\log y_{m}. 
\]%
In particular, identifying $\alpha >0$ with the vector all of whose
components are $\alpha $ (qua function constantly $\alpha $), $\langle
y^{\alpha }\rangle =(y_{1}...y_{m})^{\alpha }.$ Finally, for convenience:%
\[
\kappa _{i}\text{ or }\kappa _{m}^{i}:=p_{i}/p\qquad (i=0,1,...,m),\text{ }%
\qquad \kappa _{1}^{i}:=p_{i}/(p_{0}+p_{i})\qquad \text{(}i=1,...,m). 
\]

\noindent \textbf{Lemma 1}$_{m}$\textbf{\ (Valuation of }$Z_{1}$\textbf{\
given observation }$Y_{1}=y$\textbf{).} \textit{Put }$\kappa =(...,\kappa
_{m}^{j},...).$ \textit{Then} 
\[
\mathbb{E}[Z_{1}^{i}|Y_{1}=y,]=k^{i}\langle y^{\kappa }\rangle \text{ for }%
k^{i}=k_{m}^{i}=f^{i}\langle (1/f)^{\kappa }\rangle . 
\]

\noindent \textbf{Proof. }With the overbar notations as in Lemma 1, note
from \cite[Prop. 10.3]{GieOa} that if $\bar{T}$ has components $\bar{T}_{i}=%
\bar{X}\bar{Y}_{i},$ where $\bar{X}$ and $(...\bar{Y}_{i}...)$ are
independent, with precision parameters $\bar{p}_{i}$, then for $\kappa _{i}=%
\bar{p}_{i}/(\bar{p}_{0}+\bar{p}_{1}+...+\bar{p}_{m})$ and $\delta >0$ 
\[
\mathbb{E}[\bar{X}^{\delta }|\bar{T}=t]=K_{\delta }t_{1}^{\delta \kappa
_{1}}...t_{m}^{\delta \kappa _{m}}=K_{\delta }\langle t^{\delta \kappa
}\rangle . 
\]%
As before $\bar{X}$ and $\bar{Y}_{i}$ have conditional variances $\sigma
_{0}^{2}\Delta t$ and $\sigma _{i}^{2}\Delta t.$ Take $K_{\delta }=1$ (its
limiting value as $\Delta t\rightarrow 0;$ see Lemma 1) and note that $\bar{p%
}_{i}=(1/\sigma _{i})^{2}.$ For $\delta =\alpha _{i},$ conditioning on $%
Y_{1}=y,$ read off $k_{m}^{i}=f^{i}\langle (1/f)^{\kappa }\rangle $ from%
\[
\mathbb{E}[Z_{1}^{i}|\bar{T}]=\mathbb{E}[f^{i}X_{1}^{\delta }|\bar{T}%
]=f^{i}\langle \bar{T}^{\delta \kappa }\rangle =f^{i}\langle
(Y_{1}/f)^{1/\delta })^{\delta \kappa }\rangle =f^{i}\langle (y/f)^{\kappa
}\rangle .\qquad \square 
\]

\bigskip

\noindent \textbf{Lemma 2}$_{m}$\textbf{\ (Time-}$t$ \textbf{conditional law
of the valuation of }$Z_{1}$\textbf{, given observation }$Y_{t}$\textbf{). }%
\textit{Conditional on }$Y_{t}=y,$\textit{\ the time-}$t$ \textit{%
distribution of the time-}$1$ \textit{valuation }$\mathbb{E}[Z_{1}^{i}|Y_{1},%
\mathcal{G}_{1}]$ \textit{is that of}%
\[
k^{i}\tilde{\beta}_{m}^{i}\langle y^{\kappa }\rangle \hat{Z}_{t}^{i}, 
\]%
\textit{where, as in Lemma 1}$_{m}$ \textit{and Lemma 2:}

\noindent (i) $\kappa :=(\kappa _{m}^{1},...,\kappa _{m}^{m})$ $;$

\noindent (ii) \textit{for }$\bar{\kappa}=1-\kappa _{0}=(p-p_{0})/p,$ 
\textit{and }$\kappa _{1}^{i}=p_{i}/(p_{0}+p_{i}),$\textit{\ }%
\[
\tilde{\beta}_{m}^{i}:=\tilde{\beta}_{\text{indiv}}^{i}\cdot \tilde{\beta}_{%
\text{agg}}:\qquad \tilde{\beta}_{\text{indiv}}^{i}:=\mu (\alpha _{i},\bar{%
\kappa}\tilde{\sigma}_{0}^{2})\mu (\kappa _{1}^{i},\bar{\kappa}\tilde{\sigma}%
_{i}^{2}),\text{\quad\ }\tilde{\beta}_{\text{agg}}=\prod\nolimits_{j}\mu
(\kappa _{j},\tilde{\sigma}_{0j}^{2}); 
\]%
\textit{\ }\noindent (iii)\textit{\ }$\hat{Z}_{t}^{i}$\textit{\ is
log-normal, its underlying mean-zero normal of variance} $%
\sum\nolimits_{j}\kappa _{j}^{2}\tilde{\sigma}_{0j}^{2}$\textit{.}

\textit{In particular, this time-}$t$ \textit{distribution} \textit{has mean
given by}%
\[
\mathbb{E}[Z_{1}^{i}|Y_{t}=y,\mathcal{G}_{t}]=k_{m}^{i}\tilde{\beta}_{\text{%
indiv}}^{i}\cdot \tilde{\beta}_{\text{agg}}\langle y^{\kappa }\rangle . 
\]%
\noindent \textbf{Proof. } From Lemma 1$_{m}$ we have for $\kappa =(\kappa
_{m}^{1},...,\kappa _{m}^{j},...)$%
\[
\mathbb{E}[Z_{1}^{i}|Y_{1},\mathcal{G}_{1}]=\mathbb{E}[f^{i}X_{1}^{\alpha
_{i}}|Y_{1},\mathcal{G}_{1}]=k_{m}^{i}\langle (Y_{1})^{\kappa }\rangle . 
\]%
Conditional on $Y_{t}=(...Y_{t}^{j}...),$ by (\ref{beta-power}) of Lemma 2,
with $\delta :=\kappa _{m}^{j}$ for each $j,$ there is $\hat{Z}_{jt}=\hat{Z}%
_{jt}(\kappa ^{j})$ of unit mean and variance $\kappa _{j}^{2}\tilde{\sigma}%
_{0j}^{2},$ with 
\[
\hat{Z}_{jt}=\exp \left( \kappa _{j}[\alpha _{j}\sigma _{0}\tilde{W}%
_{1-t}^{0}+\sigma _{j}\tilde{W}_{1-t}^{j}]-\frac{1}{2}\kappa _{j}^{2}\tilde{%
\sigma}_{0j}^{2}\right) , 
\]%
such that%
\[
(Y_{1}^{j})^{\kappa _{j}}=\beta ^{j}(Y_{t}^{j})^{\kappa _{j}}\hat{Z}_{jt},%
\text{ where }\beta ^{j}=(\mu _{t}^{0}(\alpha _{j})\mu _{t}^{j})^{\kappa
_{j}}\mu (\kappa _{j},\alpha _{j}^{2}\tilde{\sigma}_{0j}^{2}), 
\]%
for $\mu _{t}^{j}:=\mu (\alpha _{j},\alpha _{j}^{2}\tilde{\sigma}_{j}^{2}).$
So by substitution%
\[
\mathbb{E}[Z_{1}^{i}|Y_{1},\mathcal{G}_{1}]=k_{m}^{i}\prod\nolimits_{j\geq
1}\beta ^{j}(Y_{t}^{j})^{\kappa _{j}}\hat{Z}_{jt}. 
\]%
Now%
\begin{eqnarray*}
\prod\nolimits_{j\geq 1}(\mu _{t}^{0}\mu _{t}^{j})^{\kappa _{j}}\mu (\kappa
_{j},\alpha _{j}^{2}\tilde{\sigma}_{0j}^{2}) &=&\prod\nolimits_{j}\mu
(\alpha _{j},\tilde{\sigma}_{0}^{2})^{\kappa _{j}}\mu (\kappa _{1}^{j},%
\tilde{\sigma}_{i}^{2})^{\kappa _{j}}\mu (\kappa _{j},\alpha _{j}^{2}\tilde{%
\sigma}_{0j}^{2}) \\
&=&\mu (\alpha _{i},\kappa \tilde{\sigma}_{0}^{2})\mu (\kappa
_{1}^{i},\kappa \tilde{\sigma}_{i}^{2})\prod\nolimits_{j}\mu (\kappa
_{j},\alpha _{j}^{2}\tilde{\sigma}_{0j}^{2}),
\end{eqnarray*}%
where $\bar{\kappa}:=\tsum\nolimits_{j\geq 1}\kappa
_{m}^{j}:=\tsum\nolimits_{j\geq 1}p_{j}/p=(p-p_{0})/p=1-\kappa _{0}.$ Put%
\begin{eqnarray*}
\hat{Z}_{t}^{i} &:&=\prod\nolimits_{j\geq 1}\hat{Z}_{jt}^{i}=\exp
\sum\nolimits_{j}\left( \kappa _{j}[\alpha _{j}\sigma _{0}\tilde{W}%
_{1-t}^{0}+\sigma _{j}\tilde{W}_{1-t}^{j}]-\frac{1}{2}\kappa _{j}^{2}\tilde{%
\sigma}_{0j}^{2}\right) \\
&=&\exp \sum\nolimits_{j\geq 1}\left( \kappa _{j}\alpha _{j}\sigma _{0}%
\tilde{W}_{1-t}^{0}-\frac{1}{2}\kappa _{j}^{2}\alpha _{j}^{2}\tilde{\sigma}%
_{0}^{2}\right) \exp \sum\nolimits_{j\geq 1}\left( \kappa _{j}\sigma _{j}%
\tilde{W}_{1-t}^{j}-\frac{1}{2}\kappa _{j}^{2}\tilde{\sigma}_{j}^{2}\right)
\end{eqnarray*}%
(since $\tilde{\sigma}_{0j}^{2}=\alpha _{j}^{2}\tilde{\sigma}_{0}^{2}+\tilde{%
\sigma}_{j}^{2}$). This is a product of independent terms each of unit
expectation. The product has variance $\sum_{j}\kappa _{j}^{2}[\tilde{\sigma}%
_{0}^{2}+\tilde{\sigma}_{j}^{2}]=\sum\nolimits_{j}\kappa _{j}^{2}\tilde{%
\sigma}_{0j}^{2}.$ We arrive at%
\[
\mathbb{E}[Z_{1}^{i}|Y_{t},\mathcal{G}_{t}]=k_{m}^{i}\beta ^{i}\langle
Y_{t}^{\kappa }\rangle \hat{Z}_{t}^{i}.\text{ }\square 
\]

\textbf{Acknowledgement. }We are most grateful to the Referee and the
Editors for their extremely detailed, scholarly and helpful report, which
has led to many improvements. We also thank our colleagues: Nick Bingham,
Dan Crisan, Lucien Foldes, Bjorn J\o rgensen, Arne L\o kka, Michael
Schroeder and David Webb, for reading and detailed commentary on an earlier
version of this paper.

\bigskip

\noindent Accounting Department, Bocconi Univesity, Milan;
Miles.Gietzmann@unibocconi.it\newline
Mathematics Department, London School of Economics, Houghton Street, London
WC2A 2AE; A.J.Ostaszewski@lse.ac.uk\newpage


\bigskip

\begin{thebibliography}{99}
\bibitem{AggE} Aggoun, L., and Elliott, R. J., \textit{Measure theory and
filtering. Introduction and applications}. Cambridge University Press, 2004.

\bibitem{BaiC} Bain, A., and Crisan, D., \textit{Fundamentals of stochastic
filtering}, Stochastic Modelling and Applied Probability 60, Springer, 2009.

\bibitem{BarNS} Barndorff-Nielsen, O. E., and Shiryaev, A., \textit{Change
of time and change of measure}. 2$^{\text{nd}}$ ed., Adv. Ser. Stat. Sci. \&
Appl. Prob. 21, World Scientific, 2015.

\bibitem{BjeS} Bjerksund P., and Stensland G., \textit{\textquotedblleft
American exchange options and a put-call transformation: a
note\textquotedblright }, J. Business Fin. \& Acc., 20.5 (1993), 761--764.

\bibitem{BinF} Bingham, N. H., and Fry, J.M., \textit{Regression: Linear
Models in Statistics,} SUMS, Springer, 2010.

\bibitem{BinK} Bingham, N. H., and Kiesel, R, \textit{Risk-neutral valuation,%
} 2$^{\text{nd}}$ ed., Springer, 2004.

\bibitem{BjoM} Bj\"{o}rk, T., and Murgoci, A., \textit{\textquotedblleft A
theory of Markovian time-inconsistent stochastic control in discrete time}%
.\textquotedblright\ Finance Stoch. 18.3 (2014), 545--592.

\bibitem{BroHM} Brody, D. C., Hughston, L. P., and Macrina, A. , \textit{%
Information-based asset pricing.} Int. J. Theor. Appl. Finance 11.1 (2008),
107--142.

\bibitem{CalN} Calzolari, A., and Nappo, G., \textit{\textquotedblleft The
filtering problem in a model with grouped data and counting observation
times\textquotedblright }, pre-print U. Rome-Sapienza, (2002).

\bibitem{CarL} Carr, P.,\ and Lee, R., \textit{\textquotedblleft Put-call
symmetry: extensions and applications\textquotedblright }, Math. Finance
19.4 (2009), 523--560.

\bibitem{CoxS} Cox, D. R., and Solomon, P. J., (2003), \textquotedblleft 
\textit{Components of variance}\textquotedblright , Chapman \& Hall.

\bibitem{DalVJ} Daley, D., and Vere-Jones, D., \textit{An introduction to
the theory of point-processes, }Vol. I (2003), Vol. II (2008), Springer.

\bibitem{DanJ} Dana, R.-A., and Jeanblanc, M., \textquotedblleft \textit{%
Financial markets in continuous time\textquotedblright }, Springer, 2003.

\bibitem{Dav1} Davis, M. H. A., \textit{Lectures on stochastic control and
nonlinear filtering.} Springer, 1984.

\bibitem{Dav2} Davis, M. H. A.,\textit{\textquotedblleft
Piecewise-deterministic Processes: a general class of non-diffusion
stochastic models\textquotedblright }, J. R. Statist. Soc. B. 46.3 (1984),
353-388.

\bibitem{Dye} Dye, R.A., \textquotedblleft \textit{Disclosure of
Nonproprietary Information}\textquotedblright , Journal of Accounting
Research 23 (1985), 123-145.

\bibitem{Fel} Feldman, D., \textquotedblleft \textit{Logarithmic
preferences, myopic decisions, and incomplete information}\textquotedblright
, J. Financial and Quantitative Analysis, 26 (1992), 619-629.

\bibitem{GieOa} Gietzmann, M., and Ostaszewski, A. J., \textquotedblleft 
\textit{Multi-firm voluntary disclosures for correlated
operations\textquotedblright }, Ann. Finance 10 (2014), no. 1, 1--45.

\bibitem{GieOb} Gietzmann, M., and Ostaszewski, A, \textquotedblleft \textit{%
Why managers with low forecast precision select high disclosure intensity:
an equilibrium analysis}\textquotedblright , Review of Quantitative Finance
and Accounting, 43 (2014), 121--153.

\bibitem{GroH} Grossman, S., and Hart, O., \textquotedblleft \textit{%
Disclosure Laws and Take-over bids\textquotedblright ,\ }Journal of Finance,
35 (1980), 323-34.

\bibitem{ImmB} Imer, O. C., and Basa, T., \textit{\textquotedblleft Optimal
estimation with limited measurements\textquotedblright }, Proc. 44th IEEE
Conference on Decision and Control, (2005),1029 - 1034, DOI:
10.1109/CDC.2005.1582293; ISBN\ 0-7803-9567-0.

\bibitem{JunK} Jung, W., and Kwon, Y., \textquotedblleft \textit{Disclosures
when the market is unsure of information endowment of
managers\textquotedblright , }Journal of Accounting Research, 26 (1988), 146
- 153.

\bibitem{KraV} Krasa, S., and Villamil, A.P.,  \textquotedblleft \textit{%
Optimal Multilateral Contracts}.\textit{\textquotedblright } Economic Theory
4.2 (1994), 167-187.

\bibitem{KenS-2} Kendall, M. G., and Stuart, A., \textquotedblleft \textit{%
The advanced theory of statistics: Vol. 2. Inference and
relationship\textquotedblright }. 4th ed., Griffin. 1979.

\bibitem{KenS-3} Kendall, M. G., and Stuart, A. \textquotedblleft \textit{%
The advanced theory of statistics: Vol. 3. Design and analysis, and
time-series\textquotedblright } 3rd ed. Hafner Press, 1976.

\bibitem{KurN} Kurtz, T., and Nappo, G., \textit{\textquotedblleft The
filtered martingale problem\textquotedblright }, in: \textit{Oxford Handbook
of Nonlinear filtering}, 129--165, Oxford Univ. Press, 2011.

\bibitem{LipS} Liptser, R.S., and Shiryaev, A. N., \textit{Statistics of
random processes}, Springer, 2001.

\bibitem{MarV} Marinovic, I., and Varas, F., \textit{The timing and
frequency of corporate disclosures}, Working paper, 2014.

\bibitem{McNFE} McNeil, A. J., Frey, R., and Embrechts, P., \textit{%
Quantitative risk management. Concepts, techniques and tools}. Princeton,
2005.

\bibitem{OstG} Ostaszewski, A., and Gietzmann, M., \textquotedblleft \textit{%
Value Creation with Dye's Disclosure Option: Optimal Risk-Shielding with an
Upper Tailed Disclosure Strategy}\textquotedblright , Review of Quantitative
Finance and Accounting, 31.1 (2008),1- 27.

\bibitem{Shi} Shin, H. S., "\textit{Disclosure Risk and Price Drift}", J.
Accounting Research, 44.2 (2006), 351-379.

\bibitem{SinAl} Sinopoloi, B., Schenata, L., Fransceschetti, M. , Poolla,
K., Jordan, M. I., and Sastry, S., \textit{\textquotedblleft Kalman
filtering with intermittent observations\textquotedblright }, IEEE Trans.
Automat. Control 49 (2004), no. 9, 1453--1464.

\bibitem{Teh} M. R. Tehranchi,\textit{\ \textquotedblleft Symmetric
martingales and symmetric smiles}\textquotedblright , Stochastic Process.
Appl. 119.10 (2009), 3785--3797.

\bibitem{Tow} Townsend, R. M., , \textquotedblleft \textit{Optimal Contracts
and Competitive Markets with Costly State Verification\textquotedblright ,}
Journal of Economic Theory 21 (1979), 265-293.
\end{thebibliography}
\end{document}